\documentclass[structabstract]{aa}  

\usepackage{natbib}

\usepackage{graphicx}
\usepackage{txfonts}

\usepackage{color}
  {}

\begin{document}

   \title{Catalogue of the morphological features in the Spitzer Survey of Stellar Structure in Galaxies (S$^4$G)
   \thanks{Tables 2 and 3 are only available in electronic form at the CDS via anonymous ftp to cdsarc.u-strasbg.fr (130.79.128.5) or via
   http://cdsweb.u-strasbg.fr/cgi-bin/qcat?J/A+A/}}


   \author{M.~Herrera-Endoqui\inst{1}
          \and 
          S.~D\'iaz-Garc\'ia\inst{1}
          \and
          E.~Laurikainen\inst{1,2}
          \and
          H.~Salo\inst{1}
         }

   \institute{Astronomy and Space Physics, University of Oulu, FI-90014, Finland\\
              \email{martin.herreraendoqui@oulu.fi}
         \and
         Finnish Centre of Astronomy with ESO (FINCA), University of Turku, V\"ais\"al\"antie 20, FI-21500, Piikki\"o, Finland}
             
\titlerunning{Catalogue of features in the S$^4$G}
\authorrunning{Herrera-Endoqui et al.}

   \date{Received 6 March 2015 / Accepted 22 June 2015}

 
  \abstract
%
{A catalogue of the features for the complete Spitzer Survey of Stellar
  Structure in Galaxies (S$^4$G), including 2352 nearby galaxies, is
  presented.  The measurements are made using  3.6 $\mu$m images,
  largely tracing the old stellar population; at this wavelength the
  effects of dust are also minimal. The measured features are the
  sizes, ellipticities, and orientations of bars, rings, ringlenses, and
  lenses. Measured in a similar manner are also barlenses (lens-like
  structures embedded in the bars), which are not lenses in the usual
  sense, being rather the more face-on counterparts of the boxy/peanut
  structures in the edge-on view. In addition, pitch angles of spiral
  arm segments are measured for those galaxies where they can be
  reliably traced. More than one pitch angle may appear for a single
  galaxy. All measurements are made in a human-supervised manner so
  that attention is paid to each galaxy.}
%
{We create a catalogue of morphological features in the complete S$^4$G.} 
{We used  isophotal analysis, unsharp masking, and fitting ellipses to measured structures.}
{We find that the sizes of the inner rings and lenses normalized to
  barlength correlate with the galaxy mass: the normalized sizes
  increase toward the less massive galaxies; it has been suggested that this is
  related to the larger dark matter content in the bar region in these
  systems. Bars in the low mass galaxies are also less concentrated,
  likely to be connected to the mass cut-off in the appearance of the
  nuclear rings and lenses. We also show observational evidence that
  barlenses indeed form part of the bar, and that a large fraction of
  the inner lenses in the non-barred galaxies could be former
  barlenses in which the thin outer bar component has dissolved.}
{}


\maketitle
%

\section{Introduction}

Galactic disks formed at redshifts z $\sim$ 1 - 2.5 when the Universe
was still clumpy and contained a large amount of gas \citep{white1978,bournaud2007,dekel2009}, in which era
galaxy encounters were also frequent. The disks formed at high
redshifts were smaller and less massive than in the local universe,
which means that a significant amount of mass was later accreted
to galaxies, and that accretion still continues at some level.  When most
of the mass accretion had terminated, galaxies started to evolve more
slowly, by star formation and by internal dynamical effects, rearranging
the mass distribution in galaxies \citep{kormendy2004}. As
imprints of this evolution, galactic disks have such morphological
structures as bars, rings, lenses, and spiral arms.  By studying these
features at mid-infrared (mid-IR) wavelengths we obtain information about the long-term
secular evolution in galaxies.

Bars can already form  at z=2 \citep{simmons2014}, but those bars are
not as frequent, and not yet similar to the bars in the nearby
universe, where even two-thirds \citep{eskridge2000,buta2015} of the
  galaxies have bars. At 3.6 $\mu$m bars appear in 55$\%$ of the
S0/a-Sc galaxies, and even in 81$\%$ of Hubble stages Scd-Sm
\citep{buta2015}.  In these very late-type galaxies bars are more
knotty, and typically the galaxies are not centrally
concentrated. Bars also evolve, for example from buckling instabilities
\citep{combes1981,pfenniger1991}, which lead to vertically thick
boxy/peanut shape bulges frequently observed in the edge-on view
\citep{jarvis1986,lutticke2000,bureau2006,yoshino2015}. 
In the more
face-on view boxy/peanut structures appear as barlenses embedded in
bars \citep{lauri2011,lauri2014,atha2014}.  By appearance alone,
barlenses are easily misinterpreted as classical bulges.

Rings appear at the resonances of bars \citep[see][]{schwarz1981,buta1996},
although other explanations for the formation of rings
have  also been suggested \citep[see][]{atha2010}. There are
nuclear, inner, and outer rings, as well as their different
varieties. In a similar manner there are also nuclear, inner, and
outer lenses \citep{kormendy1979,lauri2011}. The sizes,
shapes, and orientations of these structures tie them into the internal
dynamical evolution of galaxies \citep[see][]{buta1996}. These
structures are further connected both to visible and dark matter
content in galaxies. However, rings and lenses are not limited to
barred galaxies \citep{grouchy2010,lauri2013}, and
particularly for lenses explanations other than the resonant origin
are generally given. For example, minor mergers can create lenses
similar to those observed in the non-barred early-type galaxies
\citep{eliche2012}. Lenses are also suggested to form from disk
instabilities \citep{atha1983}, truncated star formation \citep{bosma1983},
or they might have formed via dissolution of bars into lenses
\citep{kormendy1979}.  It has also been suggested,  based on their
similarity in size, that inner lenses in the non-barred galaxies might be
former barlenses in which the outer thin bar component has dissolved
\citep{lauri2013}.

Spiral arms are generally considered as density waves propagating in
the stellar disk. Morphology of the spiral arms is connected to
  the physical properties of the galaxies. For example, pitch angle is
  found to correlate with the central mass concentration and the
  density of the HI gas in galaxies \citep{davis2015}. It also depends
   on the total galaxy mass so that the spiral arms are more open
  in galaxies with rising rotation curves, whereas those with falling
  rotation curves are generally tightly wound \citep{seigar2005,seigar2006,seigar2014}.
  The arms are more flocculent in the low mass galaxies
\citep{elmegreen1985,elmegreen2011}, where the dark matter
halos are also likely to be more dominant. Spiral arms can also be
triggered by bars or by tidal interactions \citep[see][]{kormendynorman1979,seigar2003,salo2010}.
In fact, the most prominent
grand-design spiral arms appear in interacting galaxies like M51.
Spiral arms can also participate in the formation of inner disks or
disk-like pseudobulges, both in barred and non-barred galaxies
\citep{carollo2002,boeker2004,emsellem2015}. 
Spiral arms are generally
delineated by strong star forming regions observed in the UV and
optical, but the mid-IR wavelength used in this study allows a more
reliable characterization of the spiral arms as a density response to the
long-term dynamical effects in galaxies.

In the current study we present a catalogue of the structure components
in the Spitzer Survey of Stellar Structure in Galaxies
\citep[S$^4$G;][]{sheth2010}, which contains over 2000 nearby galaxies
observed at mid-IR wavelengths.  We use the 3.6 $\mu$m images to
measure the sizes, orientations, and ellipticities of bars, rings, and
lenses. In a similar manner we also measure  barlenses, which actually
form part of the bar. For the spiral galaxies the pitch angles of the
spirals arms are also measured.  We use these measurements to discuss
the nature of rings, lenses, and barlenses in the complete S$^4$G.


\section{Sample and the database}

%
   We use the Spitzer Survey of Stellar Structure in
   Galaxies \citep[S$^4$G;][]{sheth2010}, which is a survey of 2352
   galaxies observed at 3.6 and 4.5 $\mu$m with the Infrared Array
   Camera \citep[IRAC;][]{fazio2004} on board the Spitzer Space
   Telescope. This survey is limited in volume (d$<$40 Mpc, $|$b$|$
   $>$ 30 deg), magnitude, corrected for internal extinction
   (B$_{corr}$ $<$ 15.5), and size (D$_{25}$ $>$ 1 arcmin). It covers
   all Hubble types and disk inclinations. However, as the sample was originally
   selected based on the HI 21 cm radial velocities (V$_{radio}$ $<$
   3000 km/sec) it lacks gas-poor early-type galaxies. To correct
   this bias, a supplementary sample of 465 early-type galaxies that
   fulfil the same selection criteria \citep{sheth2013} will be
   included in the sample and studied in a future work. The S$^4$G
   extends to lower galaxy luminosities than most previous samples of
   barred galaxies.

The 3.6 $\mu$m images used in this study allow a dust-free view
\citep{draine1984} of the old stellar population in galaxies.
However, although the 3.6 $\mu$m largely traces the mass in galaxies,
this wavelength is also contaminated by emission from hot dust, and by
3.3 $\mu$m emission features from polycyclic aromatic hydrocarbon
(PAH) associated  with star forming regions \citep[see][]{meidt2012}.
Although this is a drawback while estimating the mass
distribution of galaxies, it has the advantage that in the same images
we can see the star forming regions even in the locations that are
obscured by dust in the optical region. The S$^4$G images are deep,
typically reaching azimuthally averaged stellar mass surface densities
of 1M$_{\sun}$ pc$^{-2}$. The spiral arms can be followed up to 1.5
R$_{25}$ (R$_{25}$ is the radius at the optical B-band surface brightness of
25 mag arcsec$^{-2}$), which is more than in most previous optical and
near-IR studies. The pixel scale at 3.6 $\mu$m is 0.75'' and the
resolution in terms of FWHM is 2.1'' \citep[][ApJS, in press]{salo2015}.  With
this resolution not all possible nuclear bars, nuclear rings, or
nuclear lenses are visible.

The S$^4$G images are processed through pipelines, including mosaic
making of the raw data (P1), making of masks of the foreground stars (P2) and
image defects, deriving the basic photometric parameters
\citep[P3,][]{munoz2015}, and making multi-component decompositions
to the 2D flux distributions of the galaxies \citep[P4,][]{salo2015}.
We use the mosaicked 3.6 $\mu$m band images from Mu\~noz-Mateos
et al. The
estimations of the sky background levels and the orientation
parameters of the disks are from Salo et al.



\section{Measured structure components}

In this study the identification of the structure components is based
on the mean de Vaucouleurs revised Hubble-Sandage morphological
classification made by \citet{buta2015} at 3.6 $\mu$m. Concerning
the details in morphology, this is the most complete classification
done so far in the spirit of de Vaucouleurs and Sandage. It includes
the stage (E, E+, S$^0$ , S0$^0$ , S0$^+$ , S0/a, Sa, Sab, Sb, Sbc,
Sc, Scd, Sd, Sdm, Sm, Im), family (SA, SAB, SB), and variety (r, rs,
s), as well as nuclear, inner and outer rings, ringlenses, and
lenses. It also includes  other fine-structures like barlenses,
X-shaped bar morphologies, and ansae (or handles) seen at the two ends
of the bar.  Notations of the different structure components that are
considered in this study are shown in Table \ref{tab_notation}. In uncertain cases
the underscore in the classification shows what is the most likely
identification of the structure. The exact meaning of the
notations can be found in the original classification
paper by \citet{buta2015}.  Altogether the number of features
identified in the classification by Buta et al., and those that we measured  are: 
{1174/1146} bars, {805/799} rings, 294/294 lenses
and ringlenses, 90/87 nuclear structures, and 67/67 barlenses,
respectively.


\begin{table}
\small
\caption{Notation of ring and lens structures.}             
\label{tab_notation}      
\centering                          
\begin{tabular}{l l}        
\hline\hline                 
Notation & Structure  \\    
\hline                        
R & Outer ring \\
R' & Outer pseudo-ring \\
R$_1$ & Outer ring with type 1 morphology\\
R'$_1$ & Outer pseudo-ring with type 1 morphology\\
R$_2$ & Outer ring with type 2 morphology\\
R'$_2$ & Outer pseudo-ring with type 2 morphology\\
$\underline{R}$L, RL, R$\underline{L}$ & Outer ringlens\\
$\underline{R}$'L, R'L, R'$\underline{L}$ & Outer pseudo-ringlens\\
L & Outer lens\\
r & Inner ring \\
r', $\underline{r}$s, rs, r$\underline{s}$ & Inner pseudo-ring\\
$\underline{r}$l, rl, r$\underline{l}$ & Inner ringlens\\
$\underline{r}$'l, r'l, r'$\underline{l}$ & Inner pseudo-ringlens\\
l, ls, sl & Inner lens\\
bl & Barlens\\
nr & Nuclear ring\\
nr' & Nuclear pseudo-ringlens\\
nrl & Nuclear ringlens\\
nr'l & Nuclear pseudo-ringlens\\
nl & Nuclear lens\\
nl' & Nuclear lens (not fully developed) \\

Combinations: &\\

R'$_1$R'$_2$ & Outer pseudo-ring with type 1 and 2 morphology \\
R$_1$L & Outer ringlens with type 1 morphology\\
R$_1$'L & Outer pseudo-ringlens with type 1 morphology\\
R$_2$L & Outer ringlens with type 2 morphology\\
R$_2$'L & Outer pseudo-ringlens with type 2 morphology\\

\hline                                   
\end{tabular}
\end{table}
\vspace{1cm}

From the point of view of our measurements the structure components are
defined in the following manner:

\vspace{0.5cm}
{\it Rings (R, r, nr)} 
\vspace{0.5cm}

\noindent Resonance  rings are features that have well-defined
inner and outer edges. The rings can be full or non-complete
pseudo-rings, which  are further divided into different subtypes
\citep{buta2015}. Rings are located at the resonances
of bars.  Outer rings are generally associated with outer Lindblad
resonance (OLR) and typically have sizes roughly twice the size of the bar
\citep{atha1982}, depending also on the rotation curve. The OLR
subtypes are R$_1$, R$_1$', R$_2$', and R$_1$R$_2$'.  Inner rings are
associated with the inner 1/4 ultraharmonic resonance (UHR) and have sizes
similar to those of bars. Nuclear rings are located at the inner Lindblad
resonance (ILR) of the bar, and typically have sizes of hundreds of
parsecs \citep{buta1986}. In the non-barred galaxies the notation of
nuclear, inner, and outer features is based on their relative size with
respect to the size of the underlying disk. In this study we do not
study polar rings or collisional rings formed in galaxy interactions.

\vspace{0.5cm}
{\it Lenses(L, l, nl)}
\vspace{0.5cm}

\noindent Just as there are nuclear, inner, and outer
rings, there are also nuclear, inner, and outer lenses. \citet{kormendy1979}
defined lenses as structures with shallow brightness gradient
interior to a sharp outer edge, and a steep gradient thereafter. They
form part of the original classification by \citet{sandage1961} and \cite{sandage1994},
but have been systematically coded into the
classification only recently by \citet{lauri2011} and \citet{buta2015}.
Like the inner rings, the inner lenses in barred galaxies have
similar sizes to the bar, whereas the outer lenses are, by
definition, much larger than bars.

\vspace{0.5cm}
{\it Ringlenses (RL, rl, nrl)}
\vspace{0.5cm}

\noindent  Ringlenses are also divided into nuclear, inner, and outer
structures.  They are intermediate types between rings and
lenses. Ringlenses resemble rings in  that they have outer edges,
and lenses in  that the inner radius is smoother than in the
rings. 

\vspace{0.5cm}
{\it Barlenses (bl)}
\vspace{0.5cm}

\noindent Barlenses were recognized as distinct features by
\citet{lauri2011}. They were defined as lens-like structures
embedded in bars, typically having sizes of $\sim$ 50$\%$ of the bar
size, which means that they are considerably larger than the nuclear
lenses.  In appearance barlenses resemble prominent bulges, but are
different from them in  that their surface brightness profiles
drop faster at the edges.  The surface brightness profiles  along
the minor and major axes of a barlens are fairly exponential
\citep{lauri2014}. Barlenses are not lenses in the same
way as the other type of lenses, being rather 
the face-on counterparts of the vertically thick boxy/peanut structures of
bars (\citeauthor{lauri2014} 2014; \citeauthor{atha2014} 2014;
see also review by \citeauthor{laurisalo2015} 2015).

\vspace{0.5cm}
{\it Spiral arms}
\vspace{0.5cm}

\noindent There are three main types of spiral arms: grand design,
flocculent, and multi-arm \citep{elmegreen1989}. Grand design galaxies
have a well-defined two-armed spiral pattern, while flocculent galaxies
have multiple spiral arm segments. The opening of the spiral arm is
defined by the pitch angle, defined as the angle between the tangent
to the spiral arm, and the tangent to a circle in a point at some
radius from the galaxy centre.  The pitch angle is not
  necessarily constant within a galaxy. It can stay nearly constant in
  certain regions of the galaxies \citep{davis2012}, or can change
  smoothly along the galaxy radius even in grand design galaxies
  \citep{davis2015}. Possible reasons for these changes are
  discussed by \citet{davis2015}. Furthermore, the spiral arms often appear
  asymmetrically at the two sides of the galaxies \citep[see][]{elmegreen2011}.

In the current study we identify and fit the different arm segments.
Our approach takes  both the radial variations and possible
asymmetric nature of the spiral arms into account.  A detailed
morphological spiral arm classification is described by
\citet{elmegreen1989} and \citet{elmegreen2011}.

\begin{figure*}
\begin{centering}

\includegraphics[angle=0,width=9.5cm]{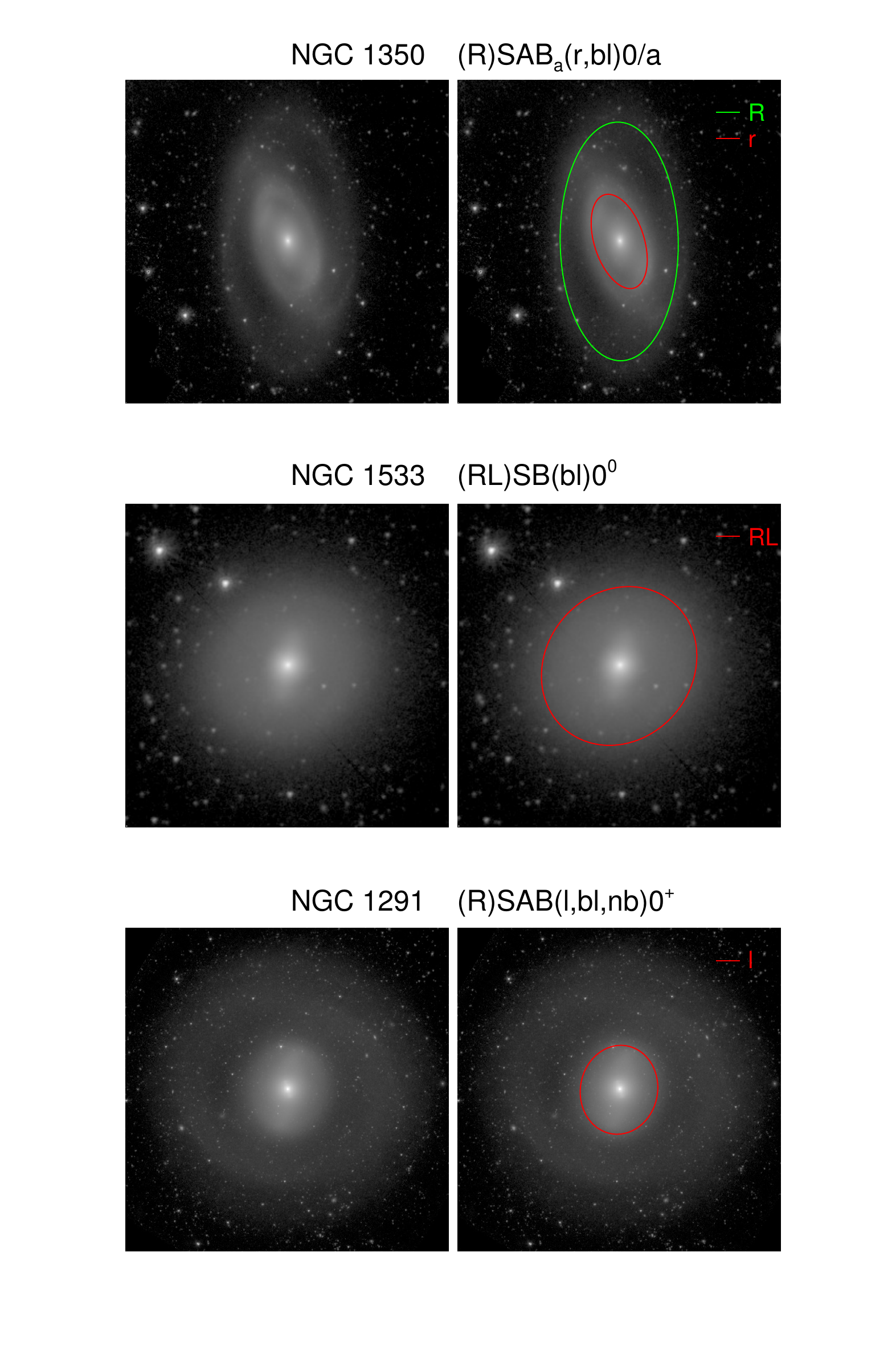}
\includegraphics[angle=0,width=9.5cm]{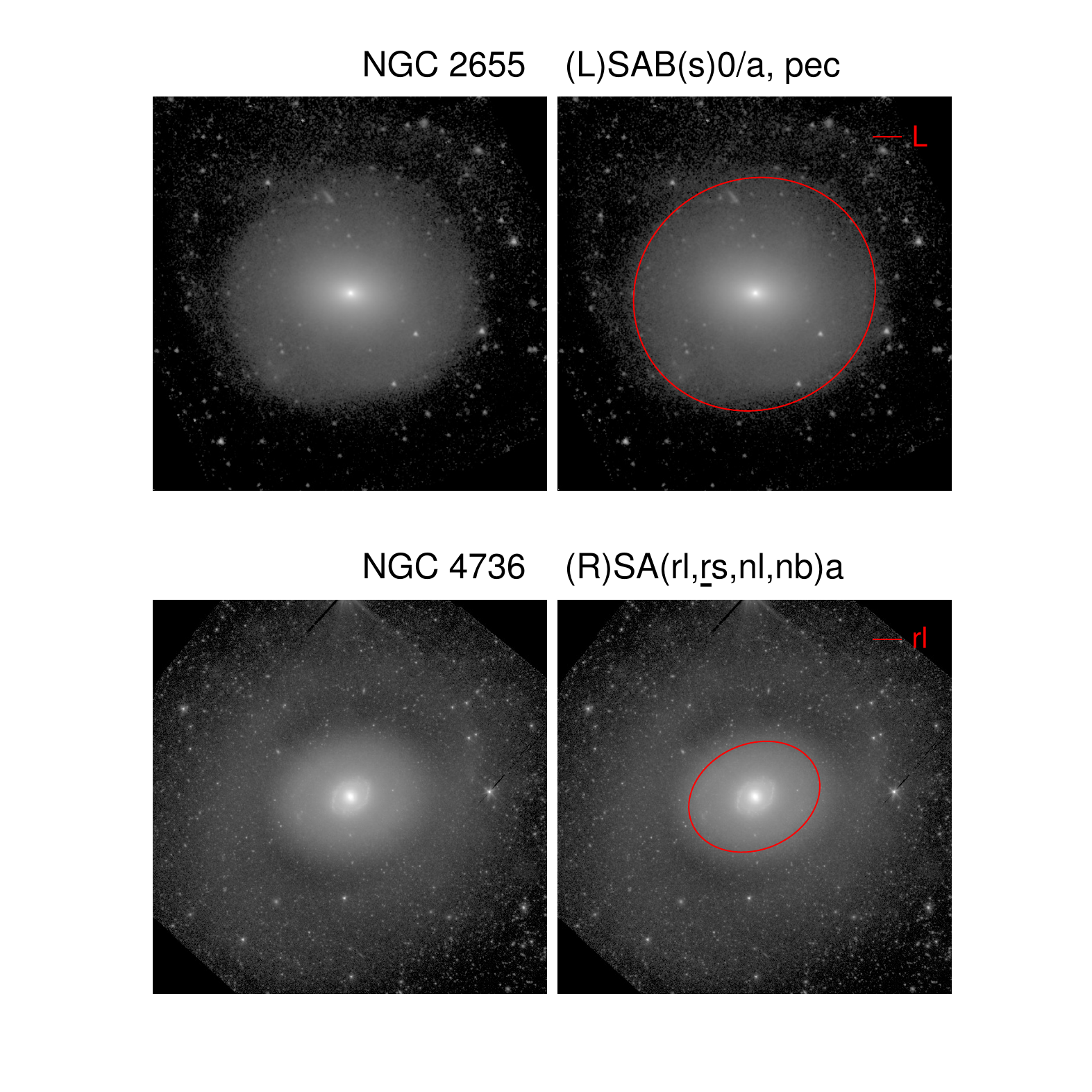}
 \caption{Examples of the structure components identified in the
classification by Buta et al. (2015). In each case one or more of
the measured features are shown on top of the 3.6 $\mu$m images
from S$^4$G.
}
\label{fig_fit_example}
%
\end{centering}
\end{figure*}

\section{Preparation of the catalogue}

In the catalogue we present the measurements of the dimensions of the
structure components identified in the classification by \citet{buta2015}.
The measured properties are the sizes, axial ratios, and
orientations of the structure components. For bars the lengths are
estimated both visually and from the isophotal ellipticity
profiles.  For spiral galaxies  the pitch angles of the measured
spiral arm segments are also given. All measurements are made with an
interactive process in which the features are first marked on top of
the images, and after that ellipses or lines are fitted to the marked 
points.  

\begin{figure}[t]
\begin{centering}
\includegraphics[angle=0,scale=.6]{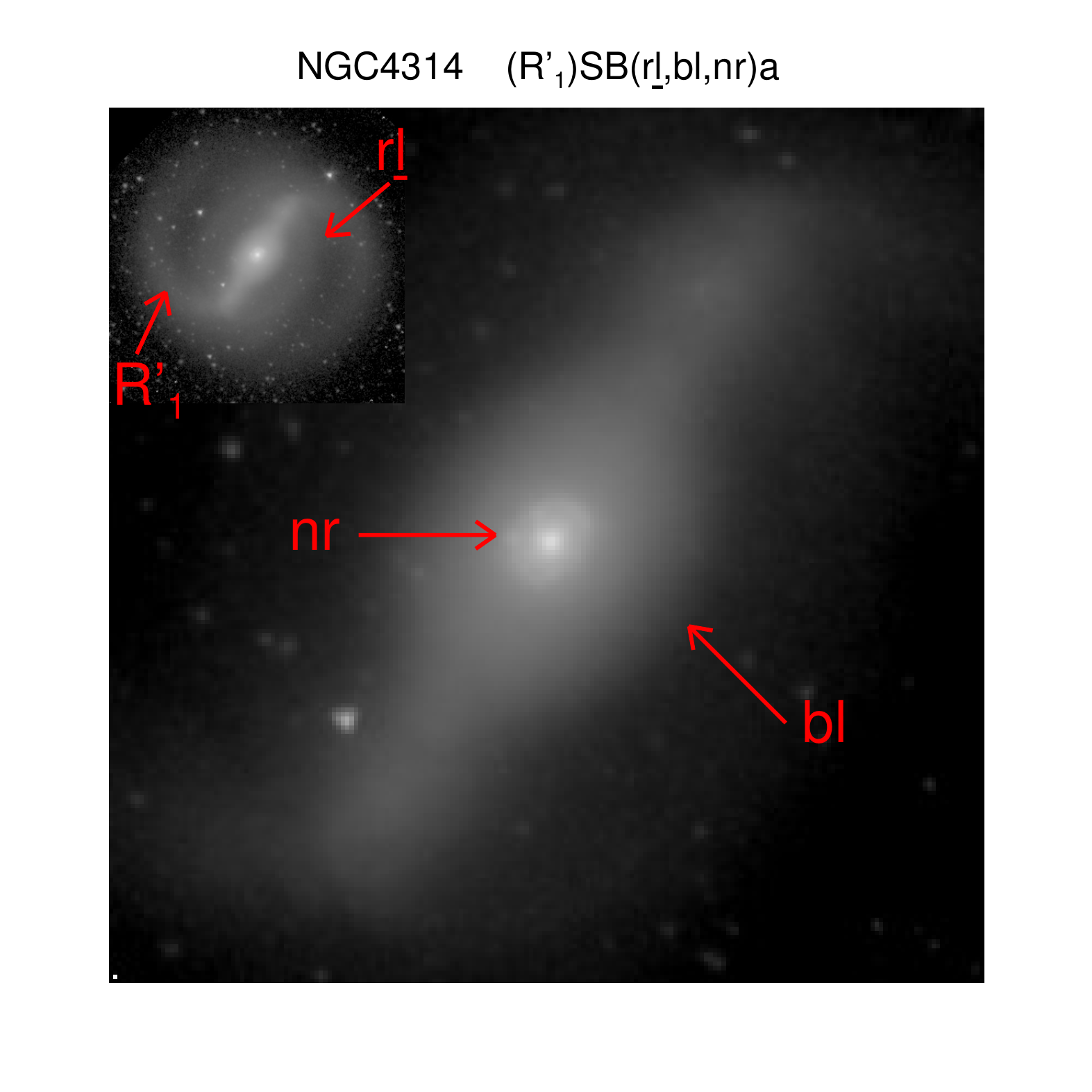} 
 \caption{NGC 4314 is a typical example of a galaxy having a barlens
   embedded inside the bar.  The image of the bar region illustrates
   the size difference between the nuclear features, compared to the
   size of a barlens. Both the star forming nuclear ring and the
   barlens are indicated by red arrows. The larger scale 3.6$\mu$m image is
   shown in the  upper left corner.}
\label{fig_bl_example}
\end{centering}
\end{figure}

Typical examples of the measured structure components are shown in
Figure \ref{fig_fit_example}, where examples of rings, ringlenses, and
lenses are given. On top of the images the size measurements of the
different components are drawn. For the sake of clarity not all
measured structures are drawn for these example galaxies.  Figure
\ref{fig_bl_example} illustrates the difference in size between the
nuclear ring and the barlens in NGC 4314. Originally, this image 
appeared in \citet{lauri2011}. Such a large difference in size between
barlenses and nuclear features is typical for all galaxies with these
structures.

\begin{figure*}
\begin{centering}
\includegraphics[angle=0,width=16cm]{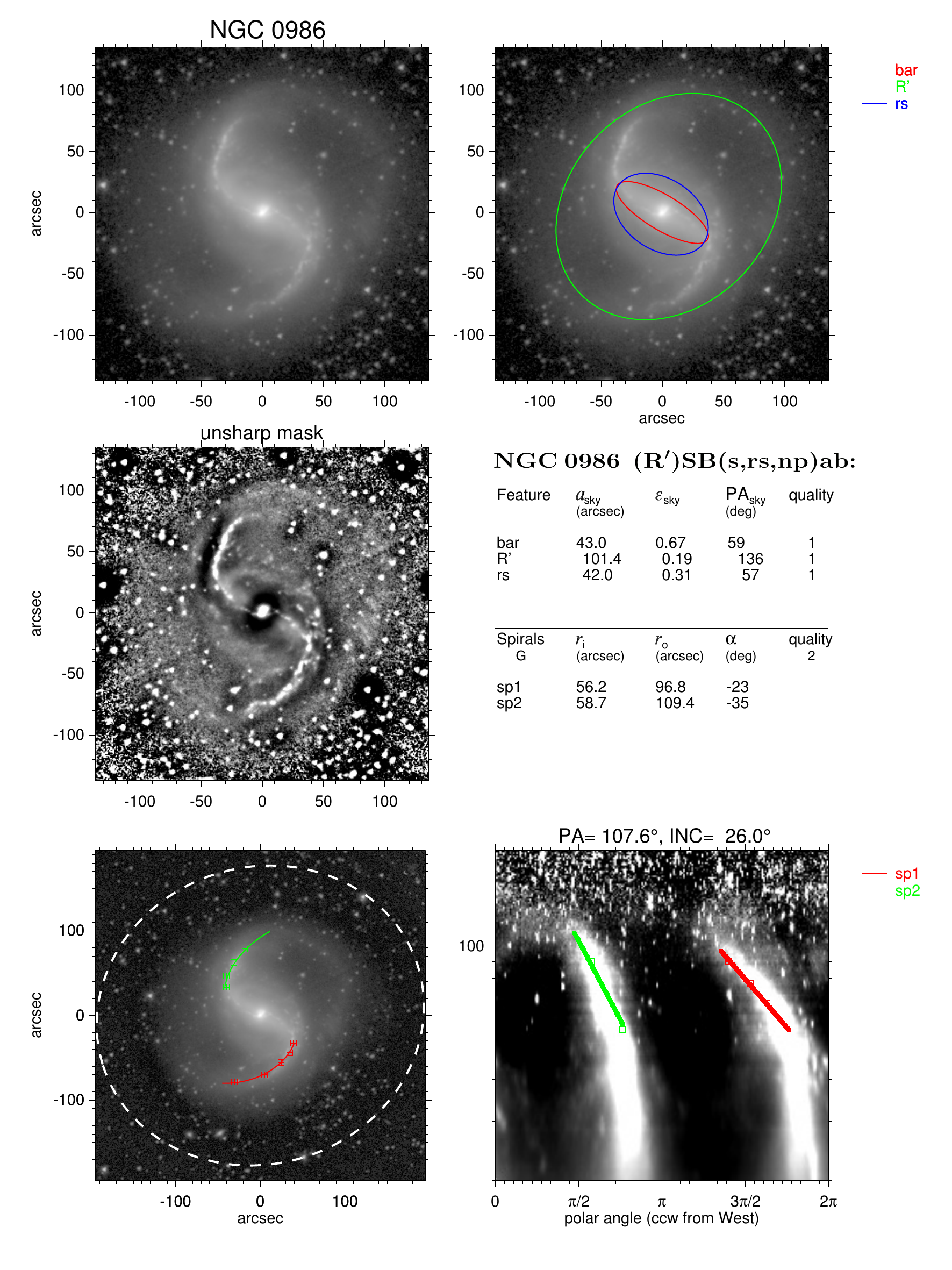}
 \caption{An example galaxy from our catalogue web page.
The panels are explained in the text.}
\label{fig_cat_example}
\end{centering}
\end{figure*}

The catalogue consists of tables (Table \ref{tab_feat} for bars, rings,
lenses, ringlenses and barlenses, and Table \ref{tab_spiral} for
spiral parameters) of the measurements for all the structure
components, and of a web page where  a page similar to
Figure \ref{fig_cat_example} appears for each galaxy\footnote{http://www.oulu.fi/astronomy/S4G\_STRUCTURES/main.html}
In Figure \ref{fig_cat_example} the two upper panels show the original
3.6 $\mu$m image: in the right panel the sizes of the measured
features are marked on top of the image. The middle panel shows an
unsharp mask image and gives the main measured parameters. The
$a_{sky}$ is the semi-major axis length, $\epsilon_{sky}$ is the
ellipticity of the structure, and $PA_{sky}$ is the position angle.  A
quality flag indicates our estimation of the reliability of the fit:
1 indicates a good fit for a feature which is unambiguously identified,
2 indicates a hard-to-trace feature (due to low
contrast with the rest of the galaxy), and 3 indicates an uncertain
feature (due to high inclination of the host galaxy
or an incomplete feature).

In the case of spiral arms, $r_i$ and $r_o$ refer to the inner and outer
radius at which the different arms where measured, and $\alpha$ is the
fitted pitch angle (positive/negative values distinguish between z- or
s-type winding in the sky). The spiral quality flag refers to our visual
estimation of the measurement quality (1.0 = good, 2.0 = acceptable). 
Highly uncertain cases in which the spiral arms are too hard to identify
were not measured.  The spiral arm classes are indicated by the letters
G, F, and M for grand design, flocculent, and multiple types,
respectively.  For a subsample of S$^4$G galaxies this classification
was done by \citet{elmegreen2011}, and for the complete sample
by \citet{buta2015}, which is also the origin of the arm classes in our
catalogue.  The lower left panel shows again the 3.6 $\mu$m image in the
sky plane, with the fitted spiral arm segments plotted on top of the
image.  The lower right panel shows the same image in logarithmic
polar coordinates after deprojection to disk plane. The assumed disk
orientation is from P4, and is indicated with the dashed ellipse in
the left panel; these values are also shown  in the title of the
right panel.

\begin{table*}
\small
\caption{Properties of bars, ring- and lens-structures in the S$^4$G.}
\label{tab_feat}
\centering
\begin{tabular}{lllccccccccc}
\hline
\hline
Galaxy & Classification & Feature  & a$_{sky}$ & PA$_{sky}$ & $\epsilon_{sky}$ & a($\epsilon$)$_{sky}$ & a$_{disk}$ & PA$_{disk}$ & $\epsilon_{disk}$   & a($\epsilon$)$_{disk}$ & Quality \\
       & (Buta et al. 2015)&       & (arcsec)  &  (deg)     &                  & (arcsec)              & (arcsec)   &   (deg)     &                     & (arcsec)               &         \\
\hline
ESO012-010 & SA$\underline{\rm B}$(s)d & bar &   14.0 &     21 &   0.50 &   12.3 &   27.3 &     43 &   0.71 &   23.1 &      1 \\
ESO012-014 & S/IB(s)m & bar &   23.8 &      5 & - & - & - & - & - & - &      3 \\
ESO013-016 & SB(rs)cd & bar &   16.1 &    166 &   0.75 &   13.8 &   16.1 &    167 &   0.61 &   13.8 &      1 \\
ESO013-016 & SB(rs)cd & rs &   24.0 &    157 &   0.25 & - &   27.8 &     90 &   0.15 & - &      1 \\
ESO015-001 & IB(s)m & bar &   14.3 &    120 & - & - & - & - & - & - &      2 \\
ESO026-001 & (R$_2^{\prime}$)SAB(s)$\underline{\rm c}$d & bar &   18.5 &     59 &   0.61 &   28.8 &   19.0 &     61 &   0.61 &   29.5 &      1 \\
ESO026-001 & (R$_2^{\prime}$)SAB(s)$\underline{\rm c}$d & R$_2^{\prime}$ &   38.4 &     54 &   0.32 & - &   39.3 &     59 &   0.31 & - &      2 \\
ESO027-001 & SA$\underline{\rm B}$(s)$\underline{\rm b}$c & bar &   13.1 &     35 &   0.53 &   18.6 &   13.8 &     45 &   0.46 &   21.4 &      1 \\
ESO027-008 & SAB(s)$\underline{\rm b}$c & bar &   14.8 &    131 & - & - & - & - & - & - &      1 \\
ESO048-017 & SB(s)m & bar &   27.9 &     58 & - & - & - & - & - & - &      3 \\
ESO054-021 & (R$^{\prime}$)SAB(s)$\underline{\rm d}$m      pec & bar &    9.0 &     69 & - & - & - & - & - & - &      2 \\
ESO054-021 & (R$^{\prime}$)SAB(s)$\underline{\rm d}$m      pec & R$^{\prime}$ &   94.3 &    102 &   0.50 & - &  114.8 &    143 &   0.31 & - &      3 \\
ESO079-005 & SB(s)m & bar &   15.6 &     13 &   0.82 &   19.1 &   17.7 &     27 &   0.76 &   21.8 &      2 \\
ESO079-007 & SB(s)dm & bar &   13.9 &     24 &   0.64 &   14.4 &   14.0 &     26 &   0.56 &   14.4 &      2 \\
ESO085-014 & SBm & bar &   23.5 &     74 & - & - & - & - & - & - &      2 \\
ESO085-047 & SB(s)m & bar &   14.8 &     43 & - & - & - & - & - & - &      2 \\
ESO107-016 & SA$\underline{\rm B}$(s)dm  sp & bar &    7.4 &     98 & - & - & - & - & - & - &      3 \\
ESO114-007 & S/IA$\underline{\rm B}$(s)m              / clump group & bar &   13.0 &     78 & - & - & - & - & - & - &      3 \\
ESO115-021 & SA$\underline{\rm B}$(s)m  sp & bar &   23.4 &     47 & - & - & - & - & - & - &      3 \\
ESO116-012 & SA$\underline{\rm B}$(s)m  sp & bar &   24.0 &     33 & - & - & - & - & - & - &      3 \\
ESO119-016 & (R$^{\prime}$)SAB(s)dm: & bar &   15.6 &     57 & - & - & - & - & - & - &      3 \\
ESO119-016 & (R$^{\prime}$)SAB(s)dm: & R$^{\prime}$ &   68.6 &     30 &   0.75 & - &   71.7 &     48 &   0.36 & - &      3 \\
ESO120-012 & IAB(s)m & bar &   24.5 &     96 & - & - & - & - & - & - &      3 \\
ESO145-025 & SAB(s)d$\underline{\rm m}$ & bar &   12.0 &    156 & - & - & - & - & - & - &      3 \\
ESO149-001 & SB(s)d  sp & bar &   17.0 &     32 & - & - & - & - & - & - &      3 \\
ESO154-023 & SB(s)$\underline{\rm d}$m  sp & bar &   46.4 &     36 & - & - & - & - & - & - &      3 \\
ESO187-035 & SB(s)m & bar &   22.1 &    116 & - & - & - & - & - & - &      3 \\
ESO187-051 & SA$\underline{\rm B}$(s)m & bar &   12.3 &     17 & - & - & - & - & - & - &      3 \\
ESO202-041 & SA$\underline{\rm B}$(r$\underline{\rm s}$)m & bar &   17.1 &    166 & - & - & - & - & - & - &      2 \\
ESO202-041 & SA$\underline{\rm B}$(r$\underline{\rm s}$)m & r$\underline{\rm s}$ &   25.3 &    173 &   0.09 & - &   34.4 &    239 &   0.27 & - &      3 \\
ESO234-043 & SA$\underline{\rm B}$(s)$\underline{\rm d}$m: & bar &   18.7 &     57 &   0.68 &   24.0 &   22.3 &     43 &   0.66 &   28.8 &      2 \\
ESO234-049 & SA(r)cd      pec & r &    8.1 &    165 &   0.58 & - &    8.8 &    165 &   0.61 & - &      1 \\
ESO236-039 & S/IABm: & bar &   10.0 &     93 &   0.62 &   14.3 &   10.9 &    119 &   0.32 &   15.9 &      2 \\
ESO237-052 & SAB(s)$\underline{\rm d}$m: & bar &   13.2 &     47 &   0.48 &   15.6 &   16.0 &      1 &   0.12 &   17.2 &      2 \\
ESO238-018 & SA$\underline{\rm B}$(s)dm & bar &    6.9 &    138 &   0.68 &    6.8 &    7.3 &    135 &   0.69 &    7.3 &      1 \\
ESO245-005 & S/IAB(s)m & bar &   50.3 &    118 & - & - & - & - & - & - &      3 \\
ESO248-002 & (L)SB(s)d & bar &   25.8 &     26 & - & - & - & - & - & - &      2 \\
ESO248-002 & (L)SB(s)d & L &   85.0 &     13 &   0.77 & - &   85.8 &    -10 &   0.06 & - &      1 \\
ESO249-035 & SB(s)d$\underline{\rm m}$  sp & bar &   20.6 &     96 & - & - & - & - & - & - &      3 \\
ESO285-048 & SB(rs)d & bar &   10.2 &    102 &   0.75 &   17.2 &   13.2 &    128 &   0.64 &   20.9 &      1 \\
ESO285-048 & SB(rs)d & rs &   21.3 &     70 &   0.64 & - &   24.5 &     39 &   0.35 & - &      2 \\
ESO287-037 & SAB(s)dm & bar &   18.7 &    163 &   0.40 &   21.1 &   20.6 &    159 &   0.45 &   22.8 &      2 \\
ESO287-043 & SA$\underline{\rm B}$(s)m  sp & bar &   18.6 &    105 & - & - & - & - & - & - &      3 \\
ESO288-013 & SA$\underline{\rm B}$(s)dm & bar &   15.9 &     35 &   0.62 &   17.2 &   16.2 &     32 &   0.61 &   17.9 &      2 \\
ESO289-026 & SAB(s)d & bar &   26.9 &    118 &   0.77 &   33.7 &   38.4 &     96 &   0.77 &   49.6 &      2 \\
ESO289-048 & SA$\underline{\rm B}$(s)d  sp & bar &   17.2 &    153 & - & - & - & - & - & - &      3 \\
ESO298-015 & SA$\underline{\rm B}$(s)$\underline{\rm d}$m & bar &   14.0 &     43 & - & - & - & - & - & - &      2 \\
ESO298-023 & S/IABm: & bar &   10.6 &     51 & - & - & - & - & - & - &      3 \\
ESO302-021 & SB(s)m  sp & bar &   16.2 &      5 & - & - & - & - & - & - &      3 \\
\hline
\end{tabular}
\tablefoot{Data for bars contains the visual estimated barlength (\textit{a}), 
the maximum ellipticity in the bar region ($\epsilon$), the visual 
estimated position angle (PA), and the barlength obtained from the
ellipticity maximum (a($\epsilon$)). They are given in both the sky
plane and the disk plane, the conversion is made using P4 orientation
parameters (Salo et al. 2015; Table 1). For bars the disk plane values
are given only when a reliable ellipticity maximum was found and the 
galaxy inclination $i<65^{\circ}$. For other features the parameters 
are obtained from fitting ellipses to points tracing the structure. 
A quality flag for our measurement is also given: 1 indicates a good
fit and unambiguously identified feature, 2 indicates a hard to trace feature, 
3 indicates an uncertain feature identification (due to high inclination of host 
galaxy or incomplete feature).}
\end{table*}


\begin{table*}
\small
\caption{Properties of spiral arms in the S$^4$G.Type of spiral arms, 
the pitch angle ($\alpha$), the inner ($r_i$), and the outer radius ($r_o$)
are given for every spiral segment (see the catalogue web page). The type of
spiral arms are taken from Buta et al. (2015): G for grand design, M for
multiple, and F for flocculent spiral arms.  Our estimation of the 
quality of the fit is also given (1.0 = good; 2.0 = acceptable).}
\label{tab_spiral}
\centering
 \begin{tabular}{llllcccc}
\hline
\hline
Galaxy & Classification     & Type               & Segment     & $\alpha$ & $r_{i}$   & r$_{o}$  & Quality  \\
       & (Buta et al. 2015) & (Buta et al. 2015) &             & (deg)    & (arcsec)  & (arcsec) &          \\
\hline
ESO 012-010 & SA$\underline{\rm B}$(s)d & F & sp1 &  -40.8 &   33.2 &   93.4 &    1.0 \\
ESO 012-010 & SA$\underline{\rm B}$(s)d & F & sp2 &  -32.2 &   37.3 &   90.4 &    1.0 \\
ESO 026-001 & (R$_2^{\prime}$)SAB(s)$\underline{\rm c}$d & M & sp1 &  -13.3 &   26.5 &   47.2 &    2.0 \\
ESO 026-001 & (R$_2^{\prime}$)SAB(s)$\underline{\rm c}$d & M & sp2 &  -12.0 &   24.9 &   49.1 &    2.0 \\
ESO 027-001 & SA$\underline{\rm B}$(s)$\underline{\rm b}$c & M & sp1 &   36.9 &   21.2 &   38.1 &    1.0 \\
ESO 027-001 & SA$\underline{\rm B}$(s)$\underline{\rm b}$c & M & sp2 &    5.7 &   39.9 &   46.4 &    1.0 \\
ESO 027-001 & SA$\underline{\rm B}$(s)$\underline{\rm b}$c & M & sp3 &   28.7 &   48.4 &  110.6 &    1.0 \\
ESO 027-001 & SA$\underline{\rm B}$(s)$\underline{\rm b}$c & M & sp4 &    2.9 &  112.4 &  120.0 &    1.0 \\
ESO 027-001 & SA$\underline{\rm B}$(s)$\underline{\rm b}$c & M & sp5 &   15.2 &   29.3 &  109.1 &    1.0 \\
ESO 027-008 & SAB(s)$\underline{\rm b}$c & G & sp1 &   17.2 &   22.5 &   67.3 &    1.0 \\
ESO 027-008 & SAB(s)$\underline{\rm b}$c & G & sp2 &   10.4 &   23.3 &   39.5 &    1.0 \\
ESO 027-008 & SAB(s)$\underline{\rm b}$c & G & sp3 &   30.1 &   41.3 &   79.4 &    1.0 \\
ESO 027-008 & SAB(s)$\underline{\rm b}$c & G & sp4 &   48.6 &   29.4 &   85.1 &    1.0 \\
ESO 054-021 & (R$^{\prime}$)SAB(s)$\underline{\rm d}$m      pec & F & sp1 &   28.9 &   26.4 &  139.1 &    2.0 \\
ESO 054-021 & (R$^{\prime}$)SAB(s)$\underline{\rm d}$m      pec & F & sp2 &   29.6 &   34.2 &  153.5 &    2.0 \\
ESO 116-012 & SA$\underline{\rm B}$(s)m  sp & F & sp1 &   -8.6 &   61.1 &   84.3 &    2.0 \\
ESO 116-012 & SA$\underline{\rm B}$(s)m  sp & F & sp2 &  -58.5 &   21.0 &   59.9 &    2.0 \\
ESO 238-018 & SA$\underline{\rm B}$(s)dm & F & sp1 &   -7.0 &   17.8 &   24.5 &    2.0 \\
ESO 238-018 & SA$\underline{\rm B}$(s)dm & F & sp2 &  -40.9 &    9.0 &   14.9 &    2.0 \\
ESO 287-037 & SAB(s)dm & F & sp1 &    8.2 &   36.6 &   47.5 &    2.0 \\
ESO 287-037 & SAB(s)dm & F & sp2 &    7.9 &   33.3 &   38.4 &    2.0 \\
ESO 288-013 & SA$\underline{\rm B}$(s)dm & F & sp1 &   -7.8 &   26.9 &   37.0 &    2.0 \\
ESO 288-013 & SA$\underline{\rm B}$(s)dm & F & sp2 &   -1.2 &   36.9 &   37.9 &    2.0 \\
ESO 288-013 & SA$\underline{\rm B}$(s)dm & F & sp3 &  -27.2 &   15.8 &   38.3 &    2.0 \\
ESO 289-026 & SAB(s)d & G & sp1 &   28.8 &   46.9 &   92.5 &    2.0 \\
ESO 289-026 & SAB(s)d & G & sp2 &   39.2 &   36.4 &   88.9 &    2.0 \\
ESO 340-042 & SB(s)$\underline{\rm d}$m & F & sp1 &  -31.8 &   20.9 &   62.3 &    2.0 \\
ESO 340-042 & SB(s)$\underline{\rm d}$m & F & sp2 &  -46.4 &   18.9 &   33.2 &    2.0 \\
ESO 340-042 & SB(s)$\underline{\rm d}$m & F & sp3 &  -22.8 &   33.8 &   59.3 &    2.0 \\
ESO 340-042 & SB(s)$\underline{\rm d}$m & F & sp4 &   -4.1 &   60.2 &   64.0 &    2.0 \\
ESO 342-050 & SA(s)bc & M & sp1 &  -32.2 &   13.0 &   30.8 &    1.0 \\
ESO 342-050 & SA(s)bc & M & sp2 &  -19.6 &   30.3 &   87.7 &    1.0 \\
ESO 342-050 & SA(s)bc & M & sp3 &  -33.6 &   20.4 &   80.8 &    1.0 \\
ESO 342-050 & SA(s)bc & M & sp4 &  -15.5 &   25.4 &   50.6 &    1.0 \\
ESO 355-026 & SA(r)bc & M & sp1 &  -11.9 &   17.0 &   49.0 &    2.0 \\
ESO 355-026 & SA(r)bc & M & sp2 &  -18.0 &   14.2 &   38.4 &    2.0 \\
ESO 355-026 & SA(r)bc & M & sp3 &  -60.9 &   19.4 &   36.3 &    2.0 \\
ESO 355-026 & SA(r)bc & M & sp4 &  -14.2 &   38.0 &   43.9 &    2.0 \\
ESO 404-027 & SAB(s)$\underline{\rm a}$b: & G & sp1 &   21.4 &   24.2 &   64.2 &    1.0 \\
ESO 404-027 & SAB(s)$\underline{\rm a}$b: & G & sp2 &    5.8 &   65.4 &   83.0 &    1.0 \\
ESO 404-027 & SAB(s)$\underline{\rm a}$b: & G & sp3 &   14.2 &   29.2 &   60.9 &    1.0 \\
ESO 407-009 & SA(s)b$\underline{\rm c}$ & M & sp1 &  -20.2 &   23.9 &   86.8 &    2.0 \\
ESO 407-009 & SA(s)b$\underline{\rm c}$ & M & sp2 &   -7.7 &   29.9 &   41.8 &    2.0 \\
ESO 440-011 & SB(rs)cd & M & sp1 &  -42.2 &   34.6 &   84.8 &    2.0 \\
ESO 440-011 & SB(rs)cd & M & sp2 &  -15.1 &   85.7 &  106.9 &    2.0 \\
ESO 440-011 & SB(rs)cd & M & sp3 &  -19.0 &   34.4 &   59.8 &    2.0 \\
ESO 443-069 & (R$^{\prime}$)SB(rs)dm & M & sp1 &   -8.3 &   36.3 &   45.4 &    2.0 \\
ESO 443-069 & (R$^{\prime}$)SB(rs)dm & M & sp2 &  -38.1 &   26.7 &   55.6 &    2.0 \\
ESO 443-069 & (R$^{\prime}$)SB(rs)dm & M & sp3 &   -7.1 &   57.5 &   80.0 &    2.0 \\
ESO 443-069 & (R$^{\prime}$)SB(rs)dm & M & sp4 &  -26.7 &   45.1 &   73.5 &    2.0 \\
ESO 443-085 & SB(s)d & M & sp1 &  -19.8 &   22.8 &   45.0 &    2.0 \\
ESO 443-085 & SB(s)d & M & sp2 &  -17.9 &   21.5 &   42.1 &    2.0 \\
ESO 485-021 & SAB(s)$\underline{\rm c}$d & G & sp1 &   50.3 &   32.5 &   66.0 &    2.0 \\
ESO 485-021 & SAB(s)$\underline{\rm c}$d & G & sp2 &   13.0 &   73.2 &   98.3 &    2.0 \\
ESO 485-021 & SAB(s)$\underline{\rm c}$d & G & sp3 &   46.5 &   32.6 &   66.7 &    2.0 \\
ESO 485-021 & SAB(s)$\underline{\rm c}$d & G & sp4 &   12.6 &   68.3 &   96.8 &    2.0 \\
\hline
\end{tabular}
\end{table*}

\subsection{Unsharp mask images}

Inspecting the unsharp mask images forms part of our process of
measuring the rings and spiral arms. Our unsharp
masks are made by  convolving the images with a Gaussian kernel, and
then dividing the original image with the convolved image. The width
of the Gaussian kernel vary from 3 to 50 pixels. In order to find the 
best way of seeing the structures in the images, the magnitude range of 
displaying the mask image is varied.
With this convolution method we avoid possible artificial
structures that might appear if the original images are simply divided
by the rebinned images. The unsharp mask images are shown for all the
S$^{4}$G galaxies in the above described catalogue pages.

\subsection{Barlengths}

\subsubsection{Measurements}

Barlengths are estimated visually for all those barred galaxies in
S$^4$G for which the end of the bar can be recognized in the
3.6 $\mu$m image, which also includes  the high inclination galaxies.
A vast majority of the galaxies
with inclinations of $i<65^{\circ}$ are further measured using the
ellipticity maximum in the bar region as a second proxy for the bar length.
The nuclear and primary bars are
measured in a similar manner. 

In the visual barlength estimation the galaxies are first displayed in
the screen with a chosen optimal brightness scale. Then one of the end
points of the bar is marked, and a line connecting the bar ends is
displayed (the bar is assumed to be symmetric with respect to the galaxy
centre). This line is interactively stretched and rotated to match the
bar in size and orientation, giving our visual barlength and position
angle estimate. { Naturally, in case of offset bars (16 galaxies)
  the  centre of the bar was not the same as the galaxy centre.
  Uncertain barlength measurements (flag=3) appear in Table \ref{tab_feat}, but
  they are excluded from all the analyses in this paper.}
Another measure of the barlength is obtained from the P4 ellipticity
profiles\footnote{The isophotal fits were recalculated for 70 barred
  galaxies in which the P4 ellipticity profiles, made with the purpose
  of determining the outer disk isophotal shape, failed to converge in
  the bar region.}
The ellipticity and position angle profiles are displayed, and the
location of the visually estimated barlength is marked on the
graphs.  The ellipticity maximum {\em \emph{associated with the bar}} is then
marked on these profiles and stored.
The barlength measurements were made by two people (MHE and SDG),
each of whom measured half of the galaxies. About 100 galaxies were
measured by both people: based on these overlapping `training'
galaxies it was checked that the two independent estimates gave 
consistent results.  With the remaining galaxies, both persons
went through all the measurements and, if necessary, the measurement was
made again after reaching an agreement on the interpretation of the bar
for the galaxy in question.

Using the ellipticity maxima for measuring barlengths is a well-known
approach \citep{wozniak1991}. In an ideal case the ellipticity
increases in the bar region, typically having a maximum at the end of
the bar, after which it suddenly drops. In the bar region the position
angle is maintained nearly constant. It has been suggested that the
ellipticity maximum gives a lower limit of barlength \citep{atha2005}.
As an alternative, the length halfway between the maximum ellipticity in
the bar region, and the first minimum after that maximum have been
used \citep{erwin2005}.  However, in S$^4$G and in galaxies more
generally, such ideal cases are not very common.  In fact, there are
many barred galaxies in which no clear drop is visible after the
maximum ellipticity. Therefore, in this study only the maximum
ellipticity is used to measure the barlength (besides the visual estimate).

Sometimes prominent spiral arms starting at the two ends of the bar
cause an additional ellipticity maximum; these spiral arms may also
extend the radial region of a nearly constant position angle. In our
approach this effect can be largely eliminated by inspecting
simultaneously the images and the radial profiles of the ellipticity
and position angle. The chosen ellipticity maximum is restricted to
the obvious bar region, and local variations in the ellipticity
profile give indications of possible spiral arms.  However, it is not
possible to eliminate the effect of the spiral arms in galaxies where,
instead of causing a separate ellipticity maximum, the spiral arms
just make the ellipticity  gradually increase up to the
distance where the spiral arms dominate.  There are also galaxies
where no clear maximum appears in the ellipticity profile. In such
cases we take a conservative approach and do not measure the length
of the bar, which  is the case mainly among the lower luminosity
galaxies at the end of the Hubble sequence.

\begin{figure*}
\begin{centering}
\includegraphics[angle=0,width=16cm]{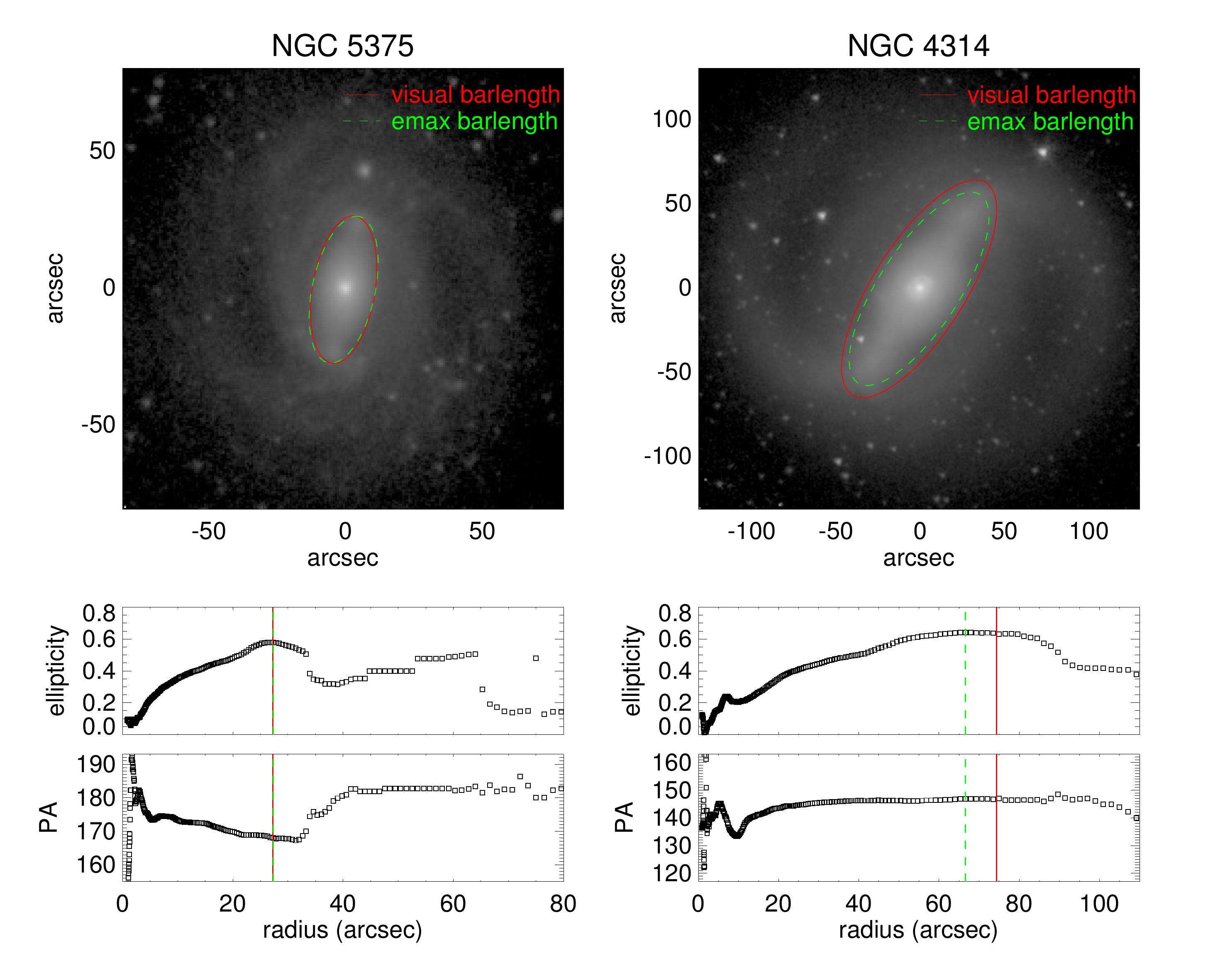}
 \caption{ Two examples (NGC 5375 and NGC 4314) of barlength
   measurements, based on visual estimation (red continuous line), and
   on ellipticity maximum (green dashed line). The bottom panels show
   the radial profiles of the ellipticity and position angle.
   Indicated with the vertical lines are the barlength measurements
   based on the two approaches.}
\label{fig_barmeas_example}
\end{centering}
\end{figure*}

Two examples of our barlength measurements are shown in Figure \ref{fig_barmeas_example}. NGC
5375 represents an ideal case where a clear ellipticity maximum
appears. The radius of this maximum gives the same barlength as
estimated visually ($r$=27''). The position angle is maintained nearly
constant to $r$=32'', and the first minimum after the ellipticity
maximum appears at $r \approx $40''.  Looking at the image it is
clear that at $r$=40'' the spiral arms are already dominating. The
position angle is maintained constant even after the ellipticity
maximum, which in this galaxy could give an upper limit of
the barlength. In NGC 4314 the bar is strong, but has no
clear maximum in its ellipticity profile, instead a broad bump appears.
The visual barlength corresponds to the radius where
  the spiral arms start on both sides of the bar. The strong spiral
  arms prevent the position angle profile from changing, which is almost
  constant up to $r=100\arcsec$. In these cases we
  looked at the image and the small variations in the ellipticity
  profile to decide the size of the bar.

Of the 1174 bars in the classification by \citet{buta2015}, we measured the
lengths for 1146 of them  visually. Of these measurements 830
were considered reliable (e.g. quality flag=1 or 2 in Table \ref{tab_feat}).
The number of barred galaxies with inclinations $i<65^{\circ}$ 
in S$^4$G is 900, of which barlengths from the ellipticity maximum
were measured for 653 galaxies. Finally,  the measurements 
were converted from the sky plane values to the
disk plane, using the P4 orientation parameters.


\begin{figure}
\begin{centering}
\includegraphics[angle=0,width=9.3cm]{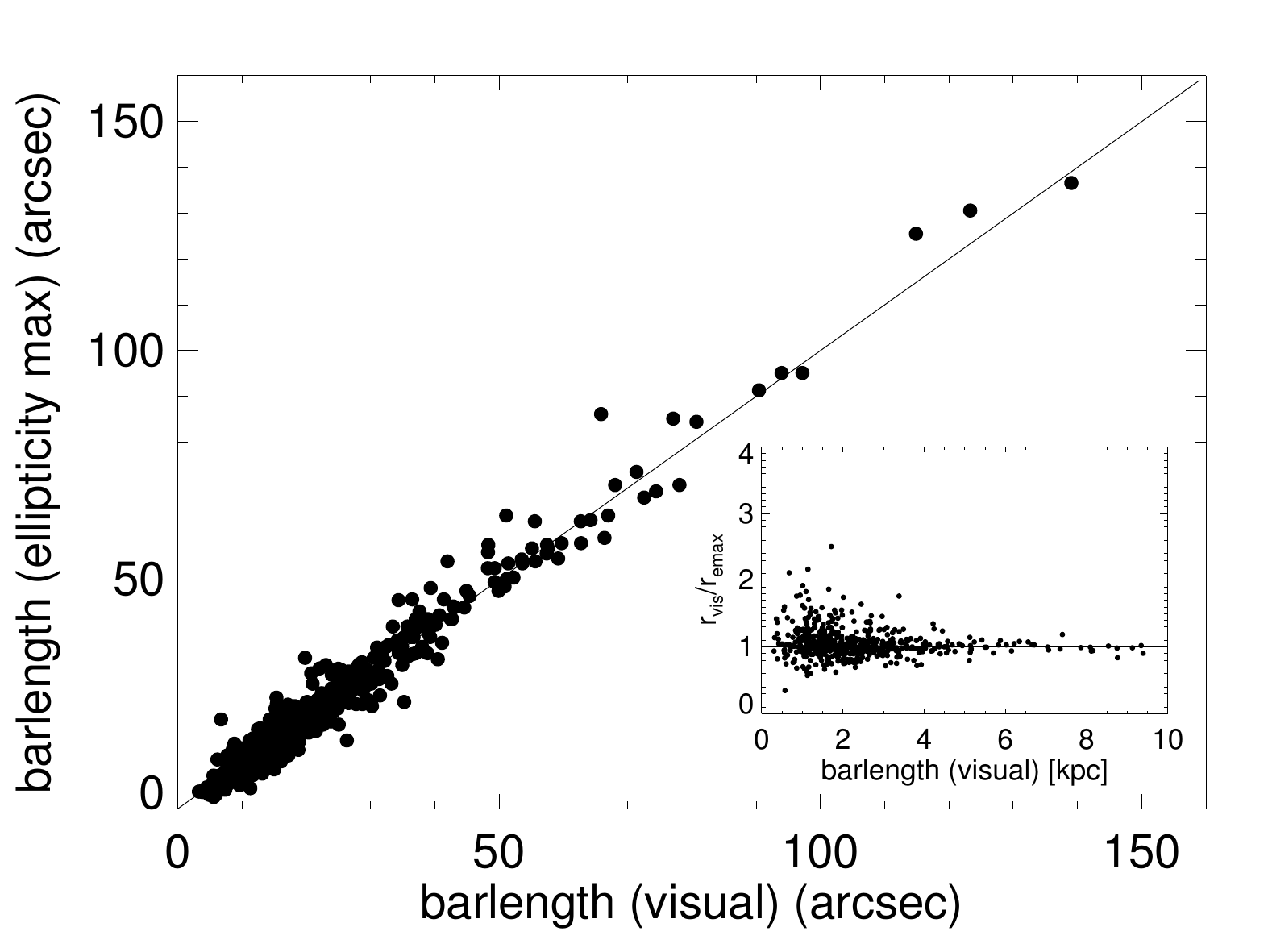} 
 \caption{A comparison of the barlength measurements, based on the two
   approaches used. {Only  galaxies with $i<65^{\circ}$ and
     with a reliable ellipticity maximum are shown.} The insert
   shows the deviations from the unit ratio.  }
\label{fig_barmeas_comp}
\end{centering}
\end{figure}

\begin{figure}
\begin{centering}
\includegraphics[angle=0,width=9.5cm]{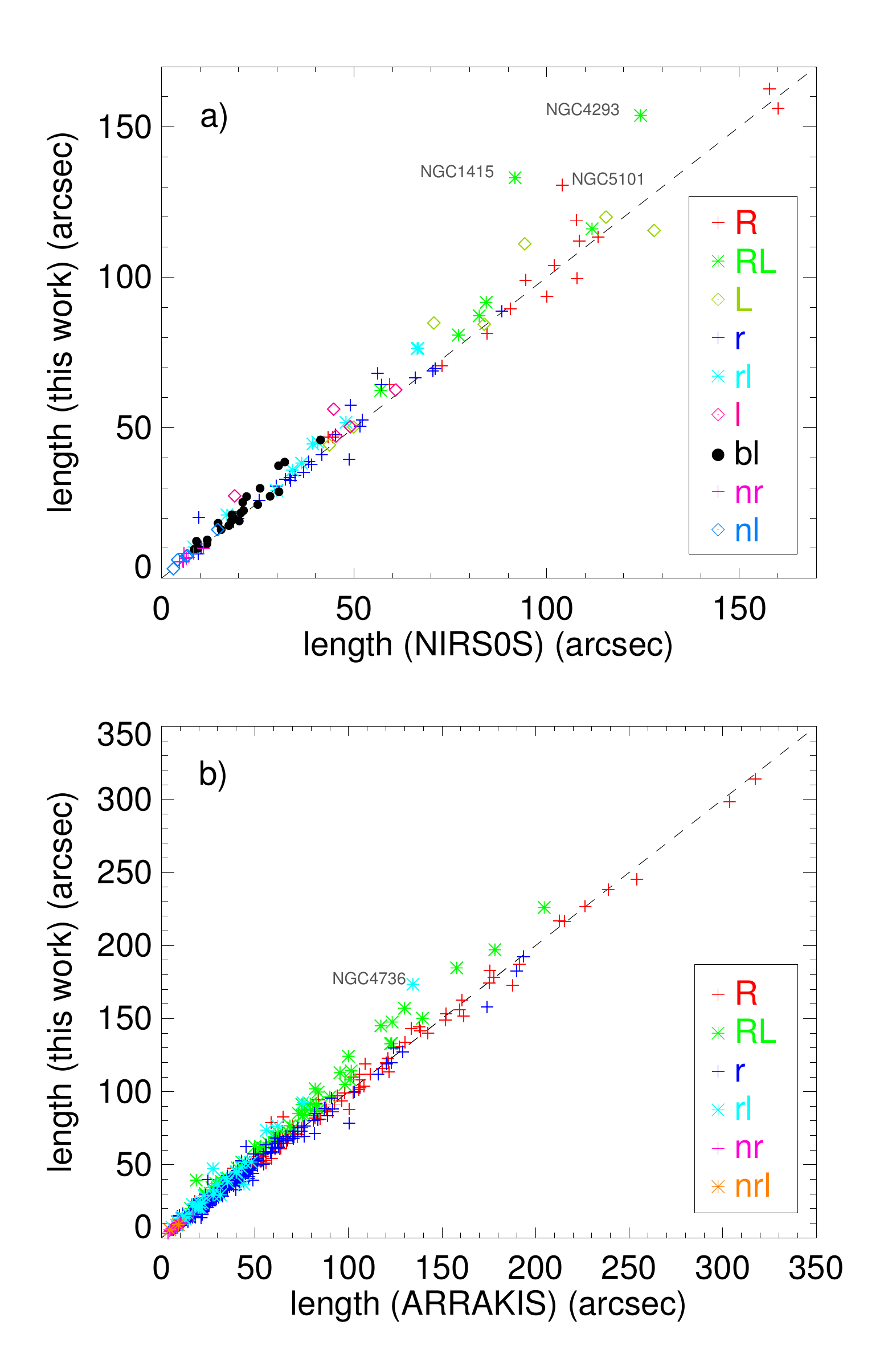} 
 \caption{Our size measurements of the different structure components
   are compared with those previously obtained for (a) the galaxies in
   NIRS0S at 2.2$\mu$m (Laurikainen et al. 2011), and (b) in ARRAKIS at 3.6$\mu$m
   (Comer\'on et al. al 2014). The galaxies for which the measurements deviate
more than 20'' are indicated by name.}
\label{fig_nirarr_comp}
\end{centering}
\end{figure}

\subsubsection{Comparison of the two barlength measurements}

A comparison of our two barlength measurements is shown in Figure \ref{fig_barmeas_comp}.
Bars with inclinations smaller than
65$^{\circ}$ and also having  a reliable ellipticity maximum in the bar region were selected.
There is some scatter in the plot, but no systematic difference
appears between the visually obtained barlengths and those measured
from the maximum ellipticity in the bar region. This is
not surprising, considering  that the galaxies were treated
individually so that both measurements generally avoid possible
problems related to spiral arms or rings at the two ends of the bar.
Dust lanes are expected to affect the measurements in the optical
region, but not in our measurements using the mid-IR images.

However, the fact that our two barlength measurements are in a fairly
good agreement with each other does not yet mean that the scatter
reflects just a random measurement uncertainty.
In fact, no single approach can give exactly the right
measurement for all bars, but we have to live with this
uncertainty. Bars have different morphologies,  from regular
elongated structures to bars with prominent ansae at the two ends of
the bar. It is clear that such bars also have  different intensity
profiles, which might slightly affect the location of the ellipticity
maximum, independent of where the bar ends.  However,  because
the deviations between our two measurements are very small, we can be
confident that the lengths of bars are measured in a consistent manner
for all galaxies in S$^4$G.


\subsection{Dimensions and ellipticities of the rings, ringlenses, and lenses}

For the measurements of the dimensions, ellipticities, and orientations
of the nuclear, inner and outer rings, ringlenses, and lenses
we use a procedure in which the image of a given galaxy is displayed
on the screen, and then the structure is visually marked.  The
brightness scale and image range are both adjusted to give optimal
visibility of the structure.
An ellipse is then analytically fitted on the marked points.
In the case of lenses and ringlenses the edges of the
structures are marked, whereas for rings the ridge-line is traced.
The fit gives  the central coordinates, the orientations, and the semi-major 
and semi-minor axis radii of the feature.
This fitting procedure is repeated three times for each structure, and the mean value is used. 
For 900 structures we also experimented with fits where the centre of
the feature was fixed to be the same as that of the galaxy. These
measurements were then compared with our original measurements, where
such a requirement was not made.  Features that were clearly
off-centred or uncertain were excluded.  We found that the mean
absolute deviations in the semi-major and semi-minor axis lengths and the
position angle were $0.48\arcsec$, $0.33\arcsec$, and $4.4^\circ$,
respectively. Clearly, for the first two parameters the deviations
are not much higher than the resolution of the images.

As rings are generally sharper than lenses or ringlenses, their
visibility in the images can be further improved using the unsharp
mask images. Therefore, instead of using the original images, rings
are identified and measured using the unsharp mask images.  In
particular, this allows us to better identify nuclear rings, which are
easily overshadowed by the luminous bulges. Also, nuclear rings are
often regions of strong star formation, which in the
optical region are easily obscured by dust. However, for the lack of
dust extinction at 3.6 $\mu$m wavelengths they are easily visible via
the PAH emission lines typical in star forming regions. 

Although barlenses are not lenses in the same sense as the other
lenses in galaxy classification, their sizes are measured in a similar manner.
By definition, barlenses appear only in barred galaxies. Furthermore,
they appear only in fairly bright galaxies, which means that they
are absent in Hubble types later than Sc.
\begin{figure*}
\begin{centering}
\includegraphics[angle=0,width=17cm]{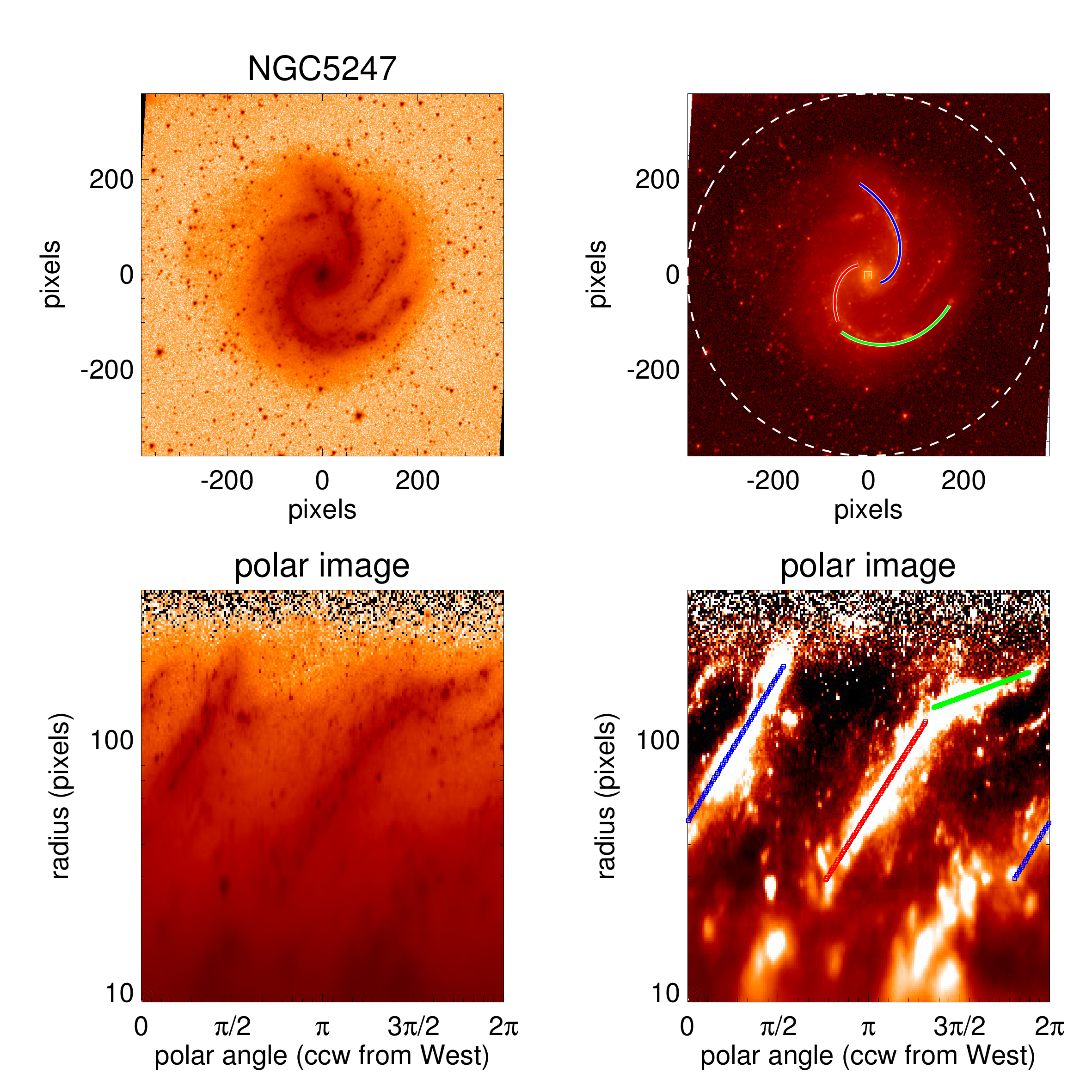} 
 \caption{An example for NGC 5247 of our measurements of the spiral
   arms.  The upper panels show the original 3.6 $\mu$m image. The
   lower left panel shows the same image in polar coordinates, and
   the lower right  panel an unsharp mask image, also in polar
   coordinates. In the right panels the locations of the measured
   spiral arms are indicated.  The different colours of the lines
   show the spiral arm segments that were fitted with different pitch angles.  }
\label{fig_spi_example}
\end{centering}
\end{figure*}

\subsection{Comparison with NIRS0S and ARRAKIS}\label{sec_comparison}


\noindent {\it Comparison with NIRS0S.} To ensure that we have measured the dimensions of
the features in a similar manner as in the previous studies, our measurements are
compared with those given in the Near-IR S0 galaxy Survey by
\citet[][hereafter NIRS0S]{lauri2011}, in which dimensions of
rings, ringlenses, lenses, and barlenses, in a sample of $\sim$200
early-type disk galaxies were measured. They used K$_s$-band images
typically reaching  surface brightnesses of 23.5 mag arcsec$^{-2}$
in the K$_s$-band, roughly equivalent to 27.5 mag arcsec$^{-2}$ in
the B-band. In S$^4$G there are 93 galaxies in common with NIRS0S. For
each galaxy we selected the structures that correspond to the same
classifications in NIRS0S and in S$^4$G, which makes 105 features in
61 galaxies (not all galaxies have rings or lenses, and not all outer
structures are identified in NIRS0S). The comparison of the sizes is
shown in Figure \ref{fig_nirarr_comp}a.

There is a good agreement between our measurements and those given in
NIRS0S, except for three measurements that differ by more than 20".  The
structures are RL in NGC1415 and NGC4293 and R in NGC5101.  In all 
three galaxies we associate the difference in size with
the different image depths in the two samples: the images are deeper
in S$^4$G thus allowing  the outer structures to  be  traced more accurately.

\vspace{0.25cm}

\noindent {\it Comparison with ARRAKIS.} Another source of reference is the atlas of resonance rings
by \citet[][hereafter ARRAKIS]{comeron2014} for a large sample of
galaxies in S$^4$G. However, in ARRAKIS lenses and barlenses were not
measured. 
Although both studies use the S$^4$G sample, the identification of the
structures in ARRAKIS and this study might vary a little because
ARRAKIS used Buta's Phase 1/Phase 2 classifications, while we use the
final classifications in \citet{buta2015}. This explains the small
difference in the number of rings in bright galaxies between the two
studies.  There is also a small difference in the way  the
measurements were done.  In ARRAKIS
residual images after subtracting the decomposition models were
used. The residual images were taken from the P4 decompositions \citep{salo2015}
in which bulges, disks, and bars were fitted with separate
functions. Ellipse fitting was then done to the residual images. This
is not much different from our approach where unsharp mask images were
used instead.  However, in principle our approach is safer because
decomposition residuals may contain artificial structures at faint
levels.

In order to compare our measurements with those made in ARRAKIS, common
galaxies between the two studies were selected, concentrating on galaxies where the
classifications were the same in both studies; this makes 1324 features
in 724 galaxies (see Figure \ref{fig_nirarr_comp}b). In general there is 
a very good agreement in the sizes: the median difference is
$1.1\arcsec$ with the mean absolute deviation of $2.8\arcsec$.
{Some differences in ringlenses (e.g. rl in NGC 4736 marked in
\ref{fig_nirarr_comp}b) can arise because we measured the structures
from the outer edges, whereas in ARRAKIS they were measured from the 
ridge-lines.}

\subsection{Pitch angles of the spiral arms}

The spiral arms are measured for all galaxies in S$^4$G, except for
the clumpy irregular galaxies at the end of the Hubble sequence.  
  The fast Fourier transform method is often used to identify spiral arms
  and to measure their pitch angles \citep{saraiva1994,seigar2005,davis2012}.
  This method  reconstructs the spiral arms in an
  efficient manner, based on the different Fourier amplitudes of
  density associated with two- or multi-armed spiral patterns.
  Drawbacks of this method are discussed by \citet{elmegreen1992}. 
  For example, the higher Fourier modes do not always
  trace real three- or four-armed spiral patterns in galaxies.  For
  asymmetric spiral arms a specific method of measuring the pitch angles
  is developed by \citet{elmegreen1992}. It is based on highlighting
  the spiral arms by making image rotation and subtracting different
  galaxy components from the original images. These methods are
  generally used in such a manner that only one pitch angle is given
  for each galaxy.

  However, we  adopted a different approach, based on visual
  inspection of the spiral arms, in the same spirit as the other
  measurements of this study. This is the first step of our analysis
of the spiral arms, allowing the identification of several pitch 
angles at different radial distances in galaxies, and taking into account
that those pitch angles might appear asymmetrical with respect to the galaxy centre.
An important part of our analysis is to display arms in polar coordinates 
which easily identifies deviations from logarithmic spirals, and therefore
also highlights the segments with distinct pitch angles in galaxies.
  
Again, the fitting procedure is interactive: the unsharp mask images
are displayed in different scales in order to see  the spiral
arms clearly, after which the spiral arms are marked.
The marked points are then displayed on top of the
deprojected galaxy image using logarithmic polar coordinates, where the 
logarithmic spirals appear as straight lines. From this
plot the points belonging to different, roughly logarithmic arm
segments are selected, and fitted with a line. The pitch angle of this
segment, together with the minimum ($r_i$) and maximum ($r_o$) radial
range of the fit are then stored.
An example of our measurements is given in Figure \ref{fig_spi_example}. The obtained pitch
angles and the radial ranges of the measurements are given in Table \ref{tab_spiral} and
the catalogue web pages. Note that we did not perform any new classification of the
spiral arms; instead the classifications by \citet{elmegreen2011} and \citet{buta2015} are shown.

\section{Analysis and discussion}

Bars identified in the classification by \citet{buta2015} in S$^4$G
are used to study the properties of bars in several papers.  The
bar fractions were discussed by Buta et al. and \citet{sheth2014}.
Barlength measurements obtained in this catalogue, combined
with bar strength measurements, are analysed  by \cite{diaz2015}.
In this study we concentrate on the rings, ringlenses, lenses, and
  barlenses, and use barlength only as a normalization factor for the
sizes of the other structure components. In the figures the visually
estimated deprojected barlengths are used.

\subsection{Resonant nature of rings, ringlenses, and lenses}

Statistics of the features identified in the classification by \citet{buta2015}
were made in the original paper. For example it was shown
that the fraction of inner rings is lowest in galaxies with the highest
bar fraction. This is the case particularly for the very
late-type galaxies (T$\geq$6), which also have low baryonic
masses and high gas fractions.  A similar tendency was also  found by
Buta et al.  for the S0s at the bright end of the mass
distribution, in which lenses largely replace the rings. A
drop in ring fraction, and an increase in lens fraction among the
early-type S0s was also recognized  by \citet{lauri2009,lauri2013}.
However, in these studies it was not clear whether the lenses could also be
interpreted as resonance structures, in a similar manner as the rings.

 An important consequence of the presence of bars in disk galaxies is 
 the appearance of dynamical resonances \citep{sellwood1993}. The disk material
 collects into these resonances giving rise to the creation of rings
 \citep[e.g.][]{schwarz1981,schwarz1984}. 
 The relevant resonances in this regard are the outer Lindblad resonance (OLR) 
 located at a radius roughly two times that of the bar, the inner ultraharmonic 
 resonance (UHR) located slightly outside the bar radius,
 and the inner Lindblad resonances (ILRs) located well inside the bar
 \footnote{For a flat rotation curve the linear treatment of resonances
 implies $R_{OLR}/R_{CR} = 1 + \sqrt2/2 \approx 1.7$, $R_{UHR}/R_{CR} = 1 - \sqrt2/4 \approx 0.65$, where
 $R_{CR}$, $R_{UHR}$, and $R_{OLR}$ are the bar corotation, UHR, and OLR distances. 
 Bars are generally assumed to end somewhat inside corotations, $1 < R_{CR}/R_{bar} < 1.7$.
 Using the nominal value $R_{CR}/R_{bar} = 1.4$ then gives $R_{OLR}/R_{bar} \sim 2.4$ and 
 $R_{UHR}/R_{bar} \sim 0.9$.}.
 As mentioned in Section 3, the outer, inner, and nuclear rings, respectively,  are believed
 to be related to these resonances.

\begin{figure*}
\begin{centering}
\includegraphics[angle=0,width=17cm]{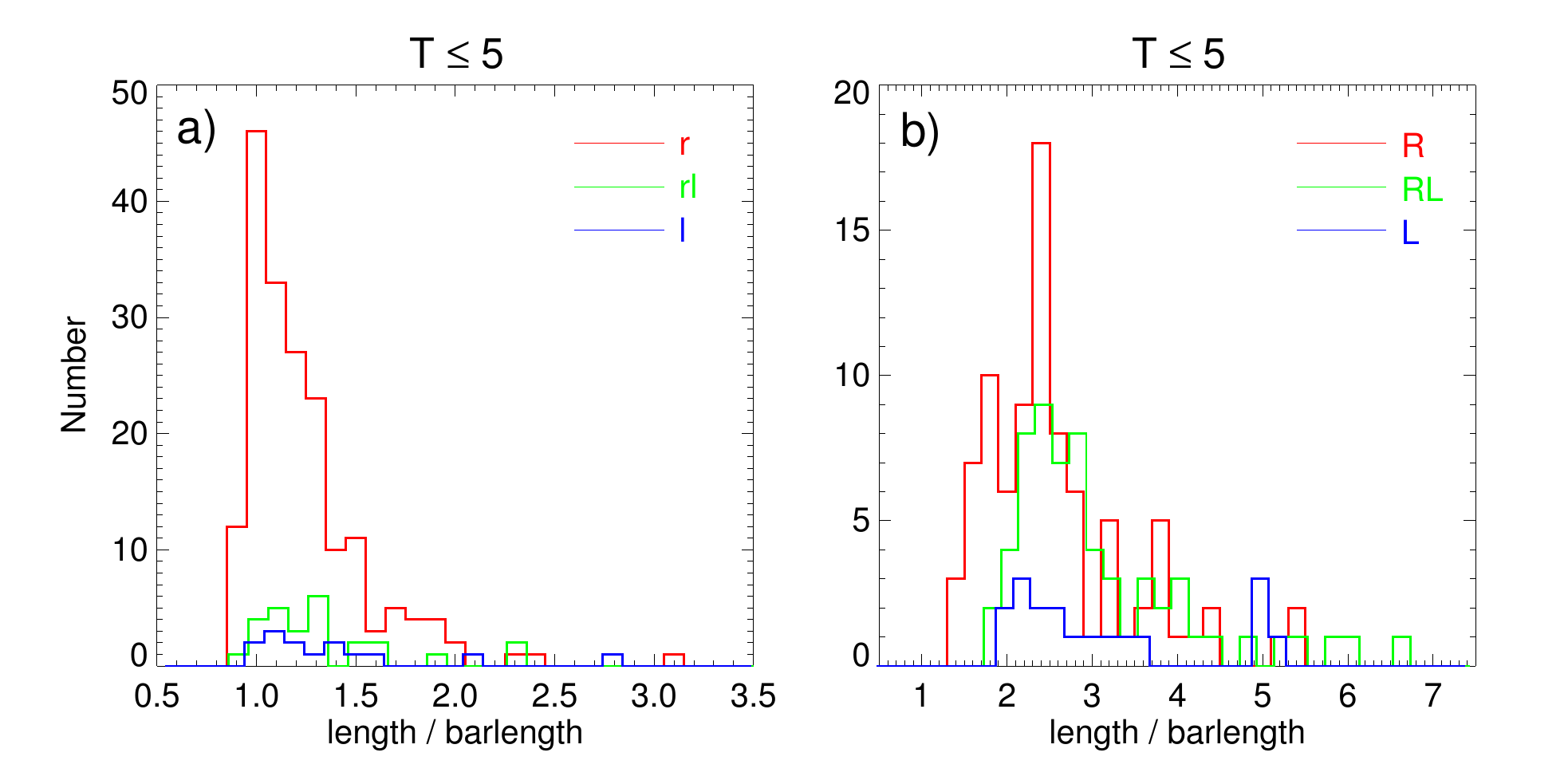} 
 \caption{The sizes of the different structure components in barred
   galaxies, normalized to the barlength. Only galaxies with Hubble
   types earlier than or equal to T=5 are shown. In (a) the inner and in (b) the
   outer features are displayed. The subtypes of all features are
   included for each type.  }
\label{fig_dist_rl_size}
\end{centering}
\end{figure*}

     In order to study the resonant nature of the inner and outer rings
     we show their sizes normalized to the size of the bar.
     This is a useful normalization as the bar corotation
     radius is always within a factor of $\sim$2 from the bar radius: 
     a theoretical minimum is R$_{cr}$/R$_{bar}$=1, since bar supporting
     orbits are not possible beyond corotation.
     These normalizations thus makes it possible to exclude features which
     definitely are not bar-related phenomena.  We concentrate on
     galaxies with Hubble stages T$\leq$5, largely covering only the bright
     galaxies that have peaked central flux
     distributions. The inner rings are peaked to length/barlength
     $\sim$1.0 (Fig. \ref{fig_dist_rl_size}a), which is similar to the rings in
     NIRS0S. There is a tail towards the higher values which, for the
     larger number of galaxies in S$^4$G, is even  clearer than in
     NIRS0S.

The inner lenses and ringlenses have very similar sizes to
the rings. For ringlenses this  confirms the result obtained in NIRS0S.
However, in NIRS0S lenses were found to be larger than the rings by a factor of
1.3. Whether this difference between NIRS0S and the current  study is
significant still needs to be confirmed by a larger sample of
lenses.  If real, it might be related to the fact NIRS0S contains gas-poor early-type S0s, which are missing in S$^4$G. The (gas-poor)
early-type S0s in NIRS0S are on average brighter than the later type
S0s and early-type spirals in the same sample \citep[see][]{lauri2011}.

The outer features are illustrated in Figure \ref{fig_dist_rl_size}b.
For the outer rings we obtain a peak at length/barlength$\approx$2.4,
again in agreement with the previous studies
\citep{kormendy1979,buta1995,lauri2013}, and also with the predictions
of the simulation models in case that the rings appear in the well-known resonances \citep[see][]{schwarz1981,buta1996}. However, the
peak is quite broad, possibly reflecting different mass distributions
of the galaxies or the fact that the $R_{CR}/R_{bar}$ ratio 
varies from galaxy to galaxy. It appears that the outer rings, ringlenses, and
lenses have very similar length distributions, most probably
indicating that they have similar physical origins.

In both panels of Figure \ref{fig_dist_rl_size} the distributions show
tails towards large relative sizes of the structures. The relative sizes larger
than 2.5 in Figure \ref{fig_dist_rl_size}a (2 cases), or larger than 5
in Figure \ref{fig_dist_rl_size}b (13 cases) are excluded from the
subsequent figures and analysis. Although they seem to be correctly
measured (flag = 1, 2 in Table \ref{tab_feat}), they are obviously not
related to the resonances of bars.

\subsection{Rings and lenses in different Hubble types and bar families}

\begin{figure*}
\begin{centering}
\includegraphics[angle=0,width=17cm]{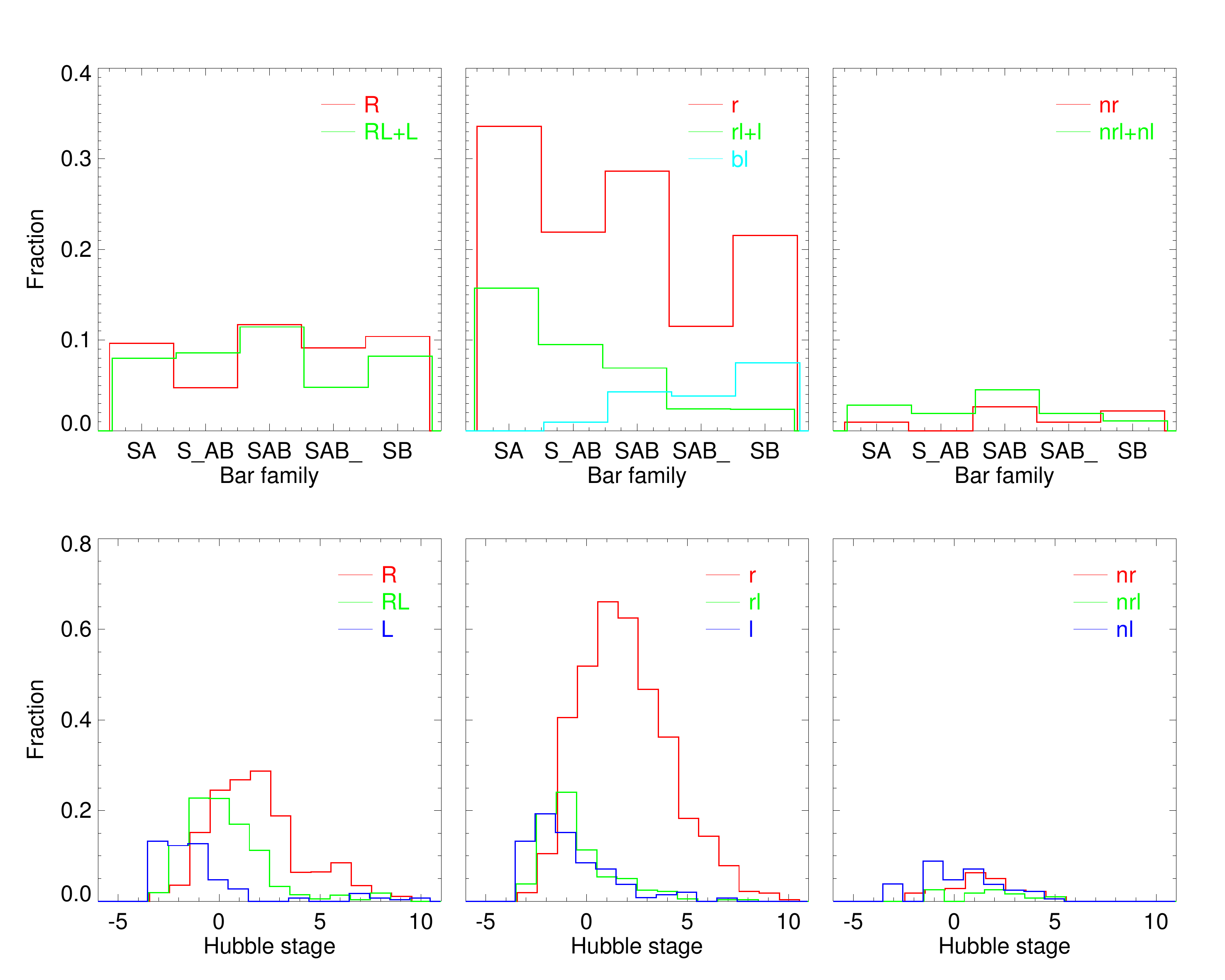} 
 \caption{Fractions of the measured features are shown as a function
   of the bar family (upper panel), and as a function of Hubble stage
   (lower panel).  The normalization is made  to all galaxies
   within each bar family class, and to the galaxies in the Hubble type
   bin. The half-integer Hubble types from Buta et al. (2015) were randomly 
   rounded to the closest smaller or larger integer value. Again, the subcategories are
   included in the features.}
\label{fig_fractions}
\end{centering}
\end{figure*}

By bar strength we mean here simply the family class of the bar.
It is generally assumed that strong bars are efficient in
redistributing matter in galaxies towards the nuclear regions
\citep{sellwood1993}, and in accumulating gas into the resonances
\citep{schwarz1981}. Curiously, it appears that weak (AB) bars rather
than strong (B) bars are more efficient in collecting matter into the
resonances where the inner rings appear, i.e. the fraction of inner
rings is clearly larger among weakly barred galaxies (see Fig.
\ref{fig_fractions}, upper middle panel).  The found tendency is even
 clearer for the inner lenses. However, it is completely lacking for
the outer and nuclear features.  This tendency for the inner
structures might be a manifestation of possible dissolution of 
bars into rings and lenses \citep[see][]{kormendy1979}. However, so
far there is no theoretical confirmation of this interpretation.
Looking at the fractions of
the features in different Hubble types, it appears that lenses and
ringlenses are concentrated in the earlier morphological types (see
Fig. \ref{fig_fractions}, lower panel). The fraction of lenses increases towards the
S0s, whereas the fraction of rings peaks at Sa Hubble types. This
is the case both for the inner, outer, and nuclear features. Similar
trends for the S0-Sa galaxies were also found  in the NIRS0S atlas.

\begin{figure*}
\begin{centering}
\includegraphics[angle=0,width=17cm]{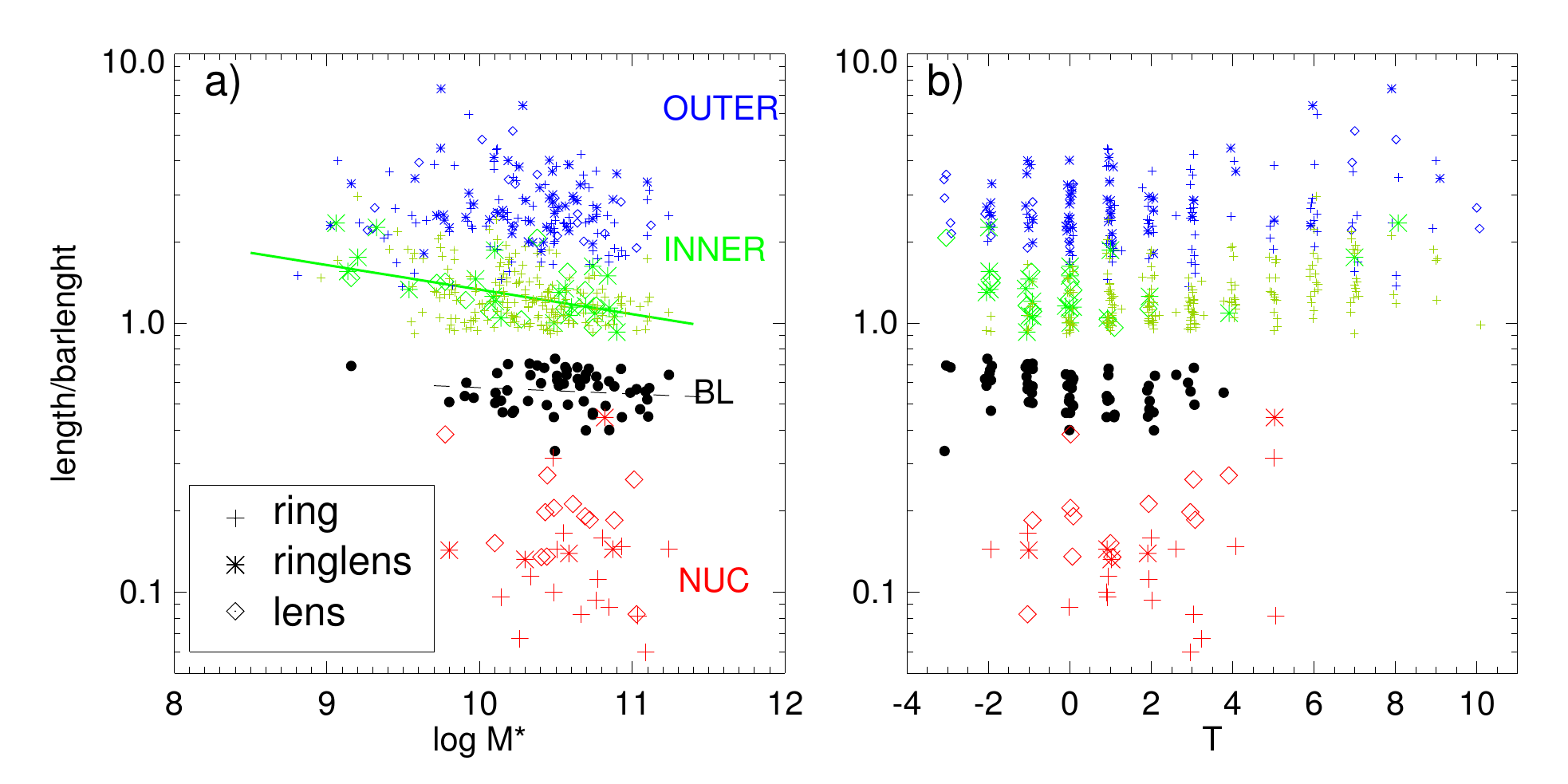} 
 \caption{The sizes of the measured features in barred galaxies
   shown as a function of (a) the parent galaxy mass and (b) the Hubble stage
   T. The sizes are normalized to the length of the bar. Shown
   with different colours are the inner (green), outer (blue), and
   nuclear (red) features, as well as barlenses (black). Within these
   feature categories rings, ringlenses, and lenses are shown with
   different symbols. Stellar masses are from P3 (Munoz-Mateos et al. 2015).
   The fit for inner structures (solid green line) is statistically significant,
   while the one for barlenses (dashed black line) is not.}
\label{fig_sizebar_mass}
\end{centering}
\end{figure*}

\subsection{Effect of the parent galaxy mass on the measured features}

\subsubsection{Sizes normalized to barlength}

The lengths of the rings, ringlenses, and lenses are shown as a
function of galaxy stellar mass and morphological type in Figure \ref{fig_sizebar_mass}. The lengths are
normalized to the size of the bar, using the measurements converted to
the disk plane.  The stellar masses are taken from \citet{munoz2013}.

A clear correlation appears between the normalized size of the inner
feature and the parent galaxy mass (Fig. \ref{fig_sizebar_mass}a,
green symbols). The shown correlation (the fitted line) is also
statistically significant, and implies that the average size of the
inner feature increases by about 50\% when the galaxy stellar mass
decreases from $10^{11} M_{\sun}$ to $10^{9} M_{\sun}$.
For the outer features (R, RL, L) the scatter is high and there is no
clear trend visible.  The scatter is particularly high for the nuclear
features (nr, nrl, nl), but an interesting characteristic for these
features is that they appear only above a parent galaxy mass $M >
10^{10} M_{\sun}$.  Since the low mass galaxies are not centrally
peaked it means that bars in those galaxies are not likely to have an
inner Lindblad resonance (ILR), which could collect gas into the
nuclear rings. Indeed, based on the structural decompositions, it has
been shown by \citet{salo2015} that the relative mass of the central
concentration in the S$^4$G galaxies rapidly drops below $10^{10}
M_{\sun}$.
This mass limit roughly corresponds to the Hubble stage T=5.  Figure
\ref{fig_sizebar_mass}b shows the sizes of the same features as a
function of Hubble type. It appears that the nuclear features and
barlenses appear only in Hubble stages T $<$ 5.

In principle an increasing mass concentration in barred galaxies is
expected to push the resonances of bars to larger radial distances
\citep{schwarz1981,combes1985,contopoulos1980}. With an
increasing mass concentration the co-rotation radius of the bar
increases, and as the bar tries to follow the co-rotation radius, 
barlength also increases. 
Consequently,  the inner and outer rings
associated with the inner 1/4 ultraharmonic (UHR) and outer Lindblad
(OLR) resonances would also be pushed further away from the galaxy
centre. Therefore, it is not immediately obvious why the normalized
sizes of the resonance features decrease towards the higher parent 
galaxy masses.

In order to better understand the observed trend of the inner ring
sizes we can look at the simulation models for IC 4214
\citep{salo1999}, where  detailed comparisons of the observed
morphological and kinematic properties of this prototypical galaxy
(with M= $5 \cdot 10^{10} M_{\sun}$) were made based on an extensive
series of simulations.  The simulations addressed the response of
sticky-particle `gas' on a rigidly rotating bar potential using
various pattern speeds and bar amplitudes.  The galaxy potential was
obtained from the $H$-band image \citep{buta1999}, and the nominal bar
amplitude $A=1$ corresponded to total potential derived directly from
the image. A value of   $A<1$ would imply that part of the axisymmetric force
field is due to a spherical dark matter halo (or additionally that the
density contrast related to the bar is less than implied by the flux
contrast). It was found (see their Table 1) that the simulated ring
sizes, for a given pattern speed, depend strongly on the bar amplitude
$A$.  For the best fitting amplitude (A=0.75) the inner ring radius
was coincident with the estimated bar length.  However, decreasing the
amplitude from $A=0.75$ to $A=0.33$ increased the inner ring radius by
about $\sim 10\%$.  Such a reduction would be at least qualitatively
consistent with the trend seen in Figure \ref{fig_sizebar_mass}, if we
assume that the decreased $A$ mimics the increased halo contribution
when moving to smaller mass galaxies. However, for the same range of
simulated $A$'s the nuclear ring increased even about $40\%$, and the
outer ring shrank by 5\%;  there are no signs of these trends in Figure
\ref{fig_sizebar_mass}. Therefore, it is likely that several factors,
like the ellipticity of the bar, pattern speed, or gas content might
all affect the trend in the inner ring size.

For example, a straightforward interpretation of larger normalized
inner ring size would be to assume that the bars of low mass systems
are slower (larger $R_{CR}/R_{bar}$) than those in high mass systems.
In this case, if we assume that the inner ring always resides at the 
UHR, the  increase by a factor of about two in the normalized inner ring size
when galaxy stellar mass M$^*$ decreases from 10$^{11}$ M$_{\sun}$ to 10$^9$ M$_{\sun}$
would imply a similar factor of two increase in $R_{CR}/R_{bar}$
(neglecting any changes in the shape of rotation curve). A similar 
increase in the normalized size of outer rings would also be expected.

\begin{figure}
\begin{centering}
\includegraphics[angle=0,width=9cm]{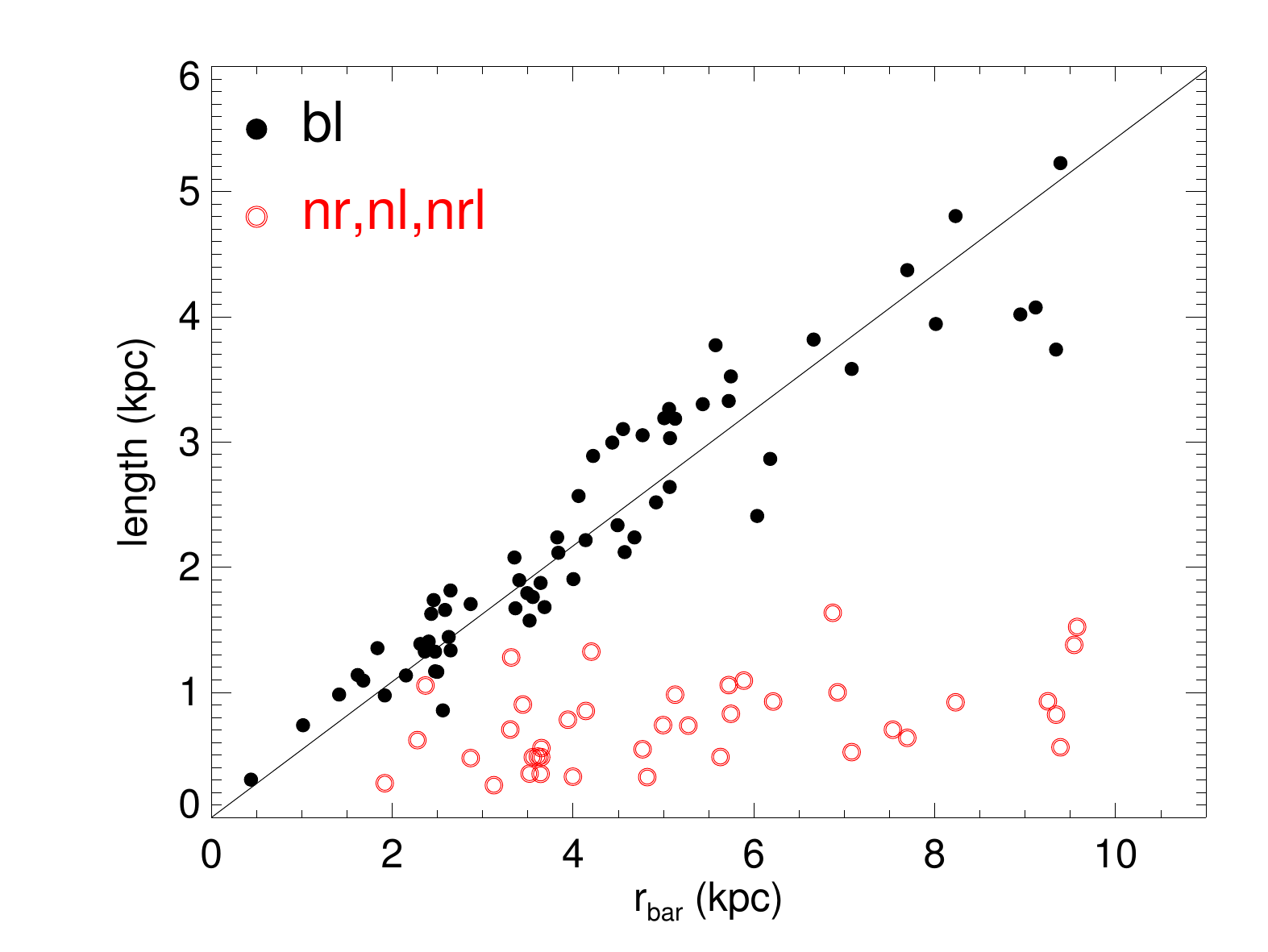} 
 \caption{The measured sizes of barlenses plotted as a function of
   the measured barlength, both given in kiloparsecs. The line shows
a linear fit between these two parameters (slope = 0.54 forcing y-intercept = 0). 
For comparison,  the measured sizes of the nuclear features are also shown, 
with no  correlation with the barlength. However, it is important to note that 
the small nuclear structures may fall below the resolution of S$^4$G images (FWHM=2.1'').
}
\label{fig_bl_bar}
\end{centering}
\end{figure}

For barlenses the relative size seems to be independent of
the parent galaxy mass (see Fig. \ref{fig_sizebar_mass}a; the fit indicated by the dashed line is not
statistically significant). 
Indeed, there is a
close linear correlation between the size of a barlens and the size of
a bar (see Fig. \ref{fig_bl_bar}). This is a manifestation of the fact that
barlenses indeed form part of the bar, which is consistent with the
idea that barlenses might indeed be the vertically thick inner parts
of bars.  For the appearance of barlenses the cut-off in the parent
galaxy mass is similar to that of the nuclear features. This is not
unexpected taking into account that barlenses appear in bright
galaxies \citep[see also][]{lauri2013}, which also have 
bright central flux concentrations.

\subsubsection{Sizes normalized to disk scalelength: comparison of barred and non-barred galaxies}

\begin{figure*}
\begin{centering}
\includegraphics[angle=0,width=19cm]{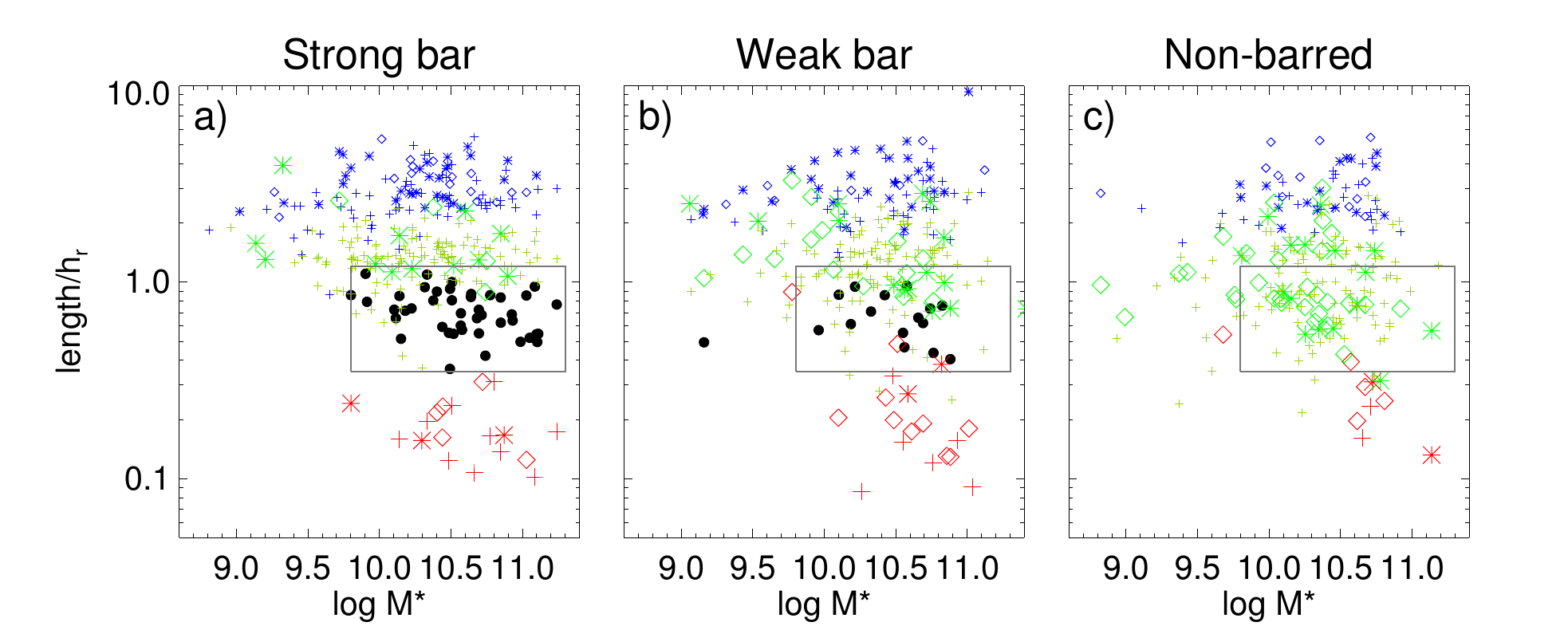} 
 \caption{A similar plot to that in Figure \ref{fig_bl_bar} (with the same symbols),
but now the normalization is to the scalelength of the disk, h$_r$, taken
from P4 (Salo et al. 2015). The three panels show the S$^4$G galaxies in different
bins of the bar family so that bar strength decreases from (a) to (c).
The box in all panels shows the region that covers the barlenses
in the barred galaxies.
}
\label{fig_sizehr_mass}
\end{centering}
\end{figure*}

In order to compare the sizes of the structure components between
barred and non-barred galaxies, the sizes need to be normalized with a
size measure independent of the bar. We have chosen to use the
scalelength of the disk (h$_r$) given by P4 decompositions
\citep{salo2015}.  The sizes of the various features as a function of
the parent galaxy stellar mass are shown in Figure
\ref{fig_sizehr_mass}, in three bins of the family classes of bars. In
\citet{buta2015} the range $T_{bar} \ge 0.75$ corresponds to SB, the
range $0.25 \leq T_{bar} < 0.75$ to SAB, and $T_{bar} < 0.25$ is for
the non-barred galaxies. It appears that for the strong bars the
inner, outer, and nuclear features occupy distinct regions,  as they do in Figure \ref{fig_sizebar_mass}a where they were
normalized to the size of the bar.  Barlenses also occupy a similar
region to that in  Figure \ref{fig_sizebar_mass}a.  However, as can be
expected the dispersion is now larger.


An interesting point in Figure \ref{fig_sizehr_mass} is that barlenses disappear while
going towards weaker bars: they seem to be gradually replaced with
rings, ringlenses, and lenses.  This is very clear while looking at
the non-barred galaxies: there is a distinct region of inner
ringlenses and lenses, covering the same area as barlenses in the
case of strongly barred galaxies.  It is worth noticing that for barred
galaxies this region is almost void of rings and lenses.
Indeed, a considerable fraction of the inner lenses in the non-barred galaxies
might be former barlenses in galaxies where the thin part of
the bar has been dissolved.  This conclusion was also made  by
\citet{lauri2013} for the early-type disk galaxies in NIRS0S,
based on the size histograms of the structures. However,
compared to NIRS0S, in
S$^4$G barlenses appear in a larger range of Hubble types and parent
galaxy masses, and the dataset is also larger.
Among the non-barred galaxies the region occupied by the
inner features in the length/h$_R$  vs. M* diagram is quite broad,
probably being a manifestation that not all inner features 
were formed by this mechanism. 

For comparison, the outer and nuclear features in Figure \ref{fig_sizehr_mass} occupy
quite similar regions in barred and non-barred galaxies. This suggests
that matter is rearranged only close to the radius of the bar, where
the orbital families behind the various structures are more easily
mixed.  If the barred potential changes during the evolution of the
galaxy it should also affect  the resonances of the
bar. However, this does not necessarily affect the already created
nuclear and outer features, if those structures, once formed, are
dynamically decoupled from the barred potential.  From Figure \ref{fig_sizehr_mass}c it
is also clear that not all inner lenses in the non-barred galaxies can
be partly dissolved bars (those with sizes much larger than barlenses
in barred galaxies). In fact, there might be two populations of
lenses. The large lenses in the mass range of $M = 10^{10} -10^{11}
M_{\sun}$ could be triggered for example by minor mergers as suggested by
\citet{eliche2012}.

\section{Summary and conclusions}

  A catalogue of the morphological features for the complete Spitzer
  Survey of Stellar Structure in Galaxies (S$^4$G), consisting of 2352
  nearby galaxies, is presented.  Using the 3.6 $\mu$m IRAC images, we
  have measured the dimensions and orientations of 1146 bars, 294
  ringlenses and lenses, 87 nuclear features, and 67 barlenses. The
  pitch angles of the spiral arms were also measured, excluding the
  messy galaxies at the end of the Hubble sequence. Multiple pitch
  angles for a single galaxy can appear.  The measured parameters are
  given in Tables \ref{tab_feat} and \ref{tab_spiral}, the complete tables being
available in electronic form. The catalogue is given as a web page\footnote{http://www.oulu.fi/astronomy/S4G\_STRUCTURES/main.html}.
Here we summarize our main conclusions:

(1) We confirm the previous results showing that inner and outer
rings are peaked to later Hubble types (T=1) than lenses and
ringlenses (T=-2). However, all types of structures appear in a wide
range of Hubble types.

(2) The inner rings and lenses are found to appear preferentially in
weakly barred (AB) and non-barred (A) galaxies.  However, the
appearance of outer and nuclear features does not depend on the family
class of the bar (see Fig. \ref{fig_fractions}).

(3) The sizes of the inner features correlate with the parent galaxy
stellar mass,  i.e. their sizes relative to the bar size
are larger in the less massive galaxies (see Fig. \ref{fig_sizebar_mass}). 

(4) The size of a barlens has a tight linear correlation with the size
of the bar (see Fig. \ref{fig_bl_bar}), which provides additional support for the conjecture
that barlenses indeed form part of the bar.

(5) Nuclear features and barlenses in barred galaxies appear only in
galaxies more massive than $\sim 10^{10} M_{\sun}$, for
barlenses  because they appear only in strong bars that appear
in bright galaxies.  On the other hand, the nuclear features are
lacking in the low mass galaxies because they lack ILRs due to low
central mass concentrations.

(6) New observational evidence is shown indicating that a large
fraction of lenses in the non-barred galaxies might be former
barlenses of bars. The outer thin part of the bar might have dissolved
or destructed, or might be too weak to be detected. This is
manifested in the length/h$_r$ vs. galaxy mass relation in such a
manner that barlenses in strong bars (B) gradually turn into inner
rings/lenses towards the non-barred galaxies (A)(see Figure
\ref{fig_sizehr_mass}).


\begin{acknowledgements}

The authors acknowledge financial support from to the DAGAL network
   from the People Programme (Marie Curie Actions) of the European
   Union's Seventh Framework Programme FP7/2007-2013 under REA grant
   agreement number PITN -GA-2011-289313. Special acknowledgement goes to
   the S4G-team (PI Kartik Sheth) for making this database available
   for us.  We also acknowledge NTT at ESO, as well as NOT and WHT in
   La Palma, where the K-band images used in this study were originally
   obtained.  EL and HS acknowledge financial support from the Academy
   of Finland. We also thank the referee for the constructive comments,
   Dr. S\'ebastien Comer\'on for very useful discussions, and MSc. Jarkko
   Laine for technical support while writing this paper.

\end{acknowledgements}

\bibliographystyle{aa} 
\bibliography{references2} 

\end{document}